\documentclass{aa}
\usepackage{txfonts}
\usepackage{graphicx}
\usepackage{natbib}
\begin{document}

\title{Counting the number of planets around GJ 581}
\subtitle{False positive rate of Bayesian signal detection methods}
\author{Mikko Tuomi\thanks{The corresponding author, \email{mikko.tuomi@utu.fi; m.tuomi@herts.ac.uk}}\inst{1,2} \and James S. Jenkins\inst{3}}

\institute{University of Hertfordshire, Centre for Astrophysics Research, Science and Technology Research Institute, College Lane, AL10 9AB, Hatfield, UK
\and University of Turku, Tuorla Observatory, Department of Physics and Astronomy, V\"ais\"al\"antie 20, FI-21500, Piikki\"o, Finland
\and Departamento de Astronom\'ia, Universidad de Chile, Camino del Observatorio 1515, Las Condes, Santiago, Chile}

\date{Received XX.XX.2012 / Accepted XX.XX.2012}

\abstract{The four-planet system aroung GJ 581 has received attention because it has been claimed that there are possibly two additional low-mass companions as well -- one of them being a planet in the middle of the stellar habitable zone.}
{We re-analyse the available HARPS and HIRES Doppler data in an attempt to determine the false positive rate of our Bayesian data analysis techniques and to count the number of Keplerian signals in the GJ 581 data.}
{We apply the common Lomb-Scargle periodograms and posterior sampling techniques in the Bayesian framework to estimate the number of signals in the radial velocities. We also analyse the HARPS velocities sequentially after each full observing period to compare the sensitivities and false positive rates of the two signal detection techniques. By relaxing the assumption that the radial velocity noise is white, we also demonstrate the consequences that noise correlations have on the obtained results and the significances of the signals.}
{According to our analyses, the number of Keplerian signals favoured by the publicly available HARPS and HIRES radial velocity data of GJ 581 is four. This result relies on the sensitivity of the Bayesian statistical analysis techniques but also depends on the assumed noise model. We also show that the radial velocity noise is actually not white and that this feature has to be accounted for when analysing radial velocities in a search for low-amplitude signals corresponding to low-mass planets.}
{Using the Bayesian analysis techniques, we did not obtain any false positives generated by noise when analysing subsets of the HARPS velocities. This suggests that any signals we detect using the Bayesian detection criteria are genuine signals present in the data. However, some such signals could be misinterpretations caused by an insufficiently accurate noise model or by (systematic) noise when independent confirmation of the signals with other data sources is not possible.}

\keywords{Methods: Statistical, Numerical -- Techniques: Radial velocities -- Stars: Individual: GJ 581}



\maketitle


\section{Introduction}

The nature of the planetary system around GJ 581 has been discussed in the literature by several authors and the preferred orbital properties and number of planets in the system have been revised whenever new data has become available. While such accumulation of knowledge is the very essence of science, contradicting results, especially those based on robust statistical arguments, have seen the system around GJ 581 become a planetary system with four established planets in its orbits and another two hypothetical ones that have been shown to exist in some studies but have been shown detections of false positives in others.

Astronomers first started counting the planets orbiting GJ 581 in 2005 when \citet{bonfils2005} reported they had found a Neptune-mass planet ($m_{p} \sin i = 16.6 $M$_{\oplus}$) with an orbital period of roughly 5.4 days. This discovery, made using the High Accuracy Radial Velocity Planet Searcher (HARPS) spectrograph \citep{mayor2003} mounted on the 3.6 m telescope at the European Southern Observatory (ESO) at La Silla, Chile, was soon accompanied by discoveries of two additional super-Earth-mass planets made by the same team in 2007 \citep{udry2007}. These two additional planets were reported to have minimum masses of 5.0 and 7.7 M$_{\oplus}$ and orbital periods of 12.9 and 84 days. \citet{udry2007} argued that the outer planet (GJ 581 d) was on a slightly eccentric orbit ($e = $0.20$\pm$0.10) but also noted that its orbit could be consistent with a circular one as was the case for the other two planets.

In 2009, \citet{mayor2009} reported that additional HARPS velocities revealed the existence of a fourth planet orbiting GJ 581 with and orbital period of 3.1 days. \citet{mayor2009} detected this planetary companion from 119 velocity measurements and claimed it to have a low minimum mass of 1.9 M$_{\oplus}$. They also revised the orbital period of GJ 581 d to 66.4 days and stated that the period of 84 days obtained by \citet{udry2007} was in fact a yearly alias of the 66.4-day signal. This is a textbook example of problems aliasing can cause to detections of periodic signals in radial velocities and other time series.

Counting the planets orbiting GJ 581 got complicated in 2010 and 2011 when \citet{vogt2010} claimed that a combined dataset with the 119 HARPS measurements of \citet{mayor2009} and another set of 122 High Resolution Echelle Spectrograph \citep[HIRES;][]{vogt1994} velocities from the 10 m Keck-II telescope at Hawaii actually contained the periodic signatures of -- not only the four planets observed by \citet{mayor2009} -- but also two other signals corresponding to low-mass planets with minimum masses of 7.0 and 3.1 M$_{\oplus}$ and orbital periods of 433 and 36.6 days, respectively. The immediate conclusion of \citet{vogt2010} was that the latter might be a good candidate for a habitable planet with its Earth-like mass and its orbit in the middle of the stellar habitable zone of GJ 581. However, the analyses of \citet{tuomi2011} and \citet{gregory2011} soon revealed that the existence of this candidate habitable world was not supported by the data and that the evidence in favour of the outer planet with an orbital period of 433 days was at most suggestive. \citet{gregory2011} and \citet{tuomi2011} concluded that there were five and four planets, respectively, in the system depending on the subjective choices of the probabilistic detection threshold and the prior densities of the parameters of the statistical models. Yet, the existence of the four planets detected by \citet{mayor2009} was easily verified using the Bayesian statistical tests of \citet{tuomi2011} and \citet{gregory2011}.

Other authors have also re-analysed the HARPS and/or HIRES data of \citet{mayor2009} and \citet{vogt2010} and received evidence that supports the existence of the hypothetical habitable zone planet GJ 581 g \citep[e.g.][]{anglada2011,dossantos2012}. However, it has been clear from the start, as also stated by almost every author discussing the system and estimating the number of planets in it, that only additional data can firmly verify or falsify the existence of the two proposed companions GJ 581 f and g.

Such data was made available by \citet{forveille2011} and increased the amount of HARPS data of GJ 581 by another 121 velocities. This increase made the total number of HARPS precision velocities 240. However, despite attempts of recovering the proposed planets f and g from this data using standard Lomb-Scargle periodograms \citep{lomb1976,scargle1982} and $\chi^{2}$ minimisations, \citet{forveille2011} concluded that these two planets could not exist orbiting the star with the masses and orbital properties reported by \citep{vogt2010}. Recently, a re-analysis of these HARPS velocities backed up by dynamical arguments was found to support the existence of a habitable-zone super-Earth with an orbital period of 32 days and a minimum mass of 2.2 M$_{\oplus}$ \citep{vogt2012} but another study taking into account the noise correlations did not find evidence in favour of the existence of this signal in the same data and also casted doubt on the significance of the signal corresponding to GJ 581 d \citep{baluev2012}. Therefore, considering the different estimates for the number of planets (parameter $k$) favoured by the data in the very recent history suggests that this debate is far from over.

If of planetary origin, any periodic variations in radial velocities should be present in data obtained using different telescope-instrument combinations. The availability of such independent confirmation is essential to exclude spurious signals that can be caused by instrument instability and insufficient noise modelling of the features in the measurement noise of the instrument.

Using the combined set of HARPS and HIRES velocities of GJ 581, we perform a global search for low-amplitude signals in an attempt to see whether the conclusions of \citet{forveille2011}, stating that $k=4$, hold for results obtained using Bayesian analysis techniques \citep[e.g.][]{tuomi2011,tuomi2012,tuomi2011b,tuomi2012b}. We also re-analyse the 240 HARPS velocities of \citet{forveille2011} to (1) enable comparison of our results with those  of \citet{bonfils2005}, \citet{udry2007}, \citet{mayor2009}, \citet{forveille2011}, \citet{vogt2012}, and \citet{baluev2011}; to (2) quantify the sensitivity of the Bayesian detection methods w.r.t. the frequentist periodogram tools used throughout the literature; and to (3) see whether the apparent sensitivity of the Bayesian detection thresholds is prone to false positives in practice. Our goal is to receive a reliable estimate for $k$ in the case of GJ 581 radial velocities and to determine the significances and properties of possible additional signals if $k > 4$. In our Bayesian analyses, we use the ``standard'' white noise models assuming independent and identically distributed measurements with Gaussian white noise but we also test models with red noise features, i.e. we model the data using a moving average (MA) model that accounts for the noise correlations in the data in a time-scale of few dozen days \citep[e.g.][]{baluev2012,tuomi2012c,tuomi2012d}.

\section{Statistical analysis}\label{sec:statistics}

\subsection{Bayesian statistics}\label{sec:bayes}

When assessing the optimal value of $k$, i.e. the number of signals statistically significantly present in a time series, we take advantage of Bayesian model selection that enables the simultaneous comparison of any number of models -- in this particular case, models with different $k$. We denote these models as $\mathcal{M}_{k}$, for all $k=0, ..., p$, and calculate the corresponding probabilities of each model in a standard way as
\begin{equation}\label{eq:model_probability}
  P(\mathcal{M}_{k} | m) = \frac{P(m | \mathcal{M}_{k}) P(\mathcal{M}_{k})}{\sum_{i=p}^{p} P(m | \mathcal{M}_{i}) P(\mathcal{M}_{i})}  ,
\end{equation}
where $m$ denotes the measurements; $P(m | \mathcal{M}_{i})$ is the Bayesian evidence, or probability of obtaining the data if the model $\mathcal{M}_{i}$ indeed was a ``correct'' model; and $P(\mathcal{M}_{i})$ is the prior probability of the model $\mathcal{M}_{i}$ that quantifies the subjective level of confidence we have on the validity of this model \emph{a priori} to analysing the measurements. As can be seen in Eq. (\ref{eq:model_probability}), the probabilities are scaled in such a way that the sum of the individual probabilities of models $\mathcal{M}_{k}$, for $k=0, ..., p$, is unity indicating that any posterior probabilities obtained using this equation are dependent on the selected set of candidate models and only reflect the \emph{relative} probabilities of the models in this set.

Finding the optimal value of $k$ based on model selections using Eq. (\ref{eq:model_probability}), however, is not as such a sufficient way of counting the periodic signals in a given data set. While a given $k$ might receive the greatest posterior probability, the claim that there really are $k$ signals in the data should not be made lightly. This is especially the case if the probability of a model with some lower value of $k$ receives estimates of similar magnitude. For this reason, we require that the probability of a model $\mathcal{M}_{k+1}$ has to be at least 150 times that of a model $\mathcal{M}_{k}$ to conclude that there is evidence in favour of the existence of $k+1$ Keplerian signals in the data. This threshold corresponds to an interpretation of ``strong evidence'' given by \citet{kass1995} to make decisions based on data. While completely subjective, we adopt this threshold but do not consider signals to exist in the data unless the other two criteria of \citet{tuomi2012} are also satisfied, i.e. that the estimates of the velocity amplitude parameter $K$ are statistically significantly greater than zero and that the period $P$ is constrained from above and below. If all these criteria are satisfied, we state that the existence of a signal is supported by data.

While the prior probabilities in Eq. (\ref{eq:model_probability}) are only simple numbers, the Bayesian evidences are in fact complicated integrals over the whole parameter space of model parameters that depend on the value of the product of a likelihood function and a prior probability density of the model parameters, i.e. the formulation of the statistical model and the physical interpretation of its parameters given the available data. For a given model, this integral, sometimes called the marginal integral, is of essence in any model comparison problems in the Bayesian context and the ability to compare models in a probabilistic way depends on the ability to estimate this integral. We write this integral as
\begin{equation}\label{eq:integral}
  P(m | \mathcal{M}_{k}) = \int_{\theta_{k} \in \Omega_{k}} l(m | \theta_{k}, \mathcal{M}_{k}) \pi(\theta_{k} | \mathcal{M}_{k}) d \theta_{k} ,
\end{equation}
where the parameters of the $k$th model, $\theta_{k}$, in the parameter space $\Omega_{k}$, are interpreted using the likelihood function $l$ and their prior probability density $\pi$ that describe the mathematical details as well as the physical interpretation of the statistical model $\mathcal{M}_{k}$. This is also the reason we write the likelihood and the prior as conditioned on the model.

We estimated the integral in Eq. (\ref{eq:integral}) using the truncated posterior mixture (TPM) estimate of \citet{tuomi2012b}. This estimation requires the availability of a statistically representative sample from the parameter posterior density and we drew such samples using the adaptive Metropolis algorithm \citep{haario2001} that is simply an adaptive version of the famous Metropolis-Hastings Markov chain Monte Carlo (MCMC) algorithm \citep{metropolis1953,hastings1970}. This sampling technique has been used succesfully in e.g. \citet{tuomi2012}.

Apart from the chosen set of candidate models (i.e. the definition of their likelihood functions), the only other subjective issue in Bayesian statistics is the prior information quantified using prior probability densities of the model parameters and prior probabilities of the different models. Standard (frequentist) analyses of radial velocity data actually incorporate prior information as well, which makes them somewhat Bayesian in reality, but do not easily enable testing different prior opinions in practice. For instance, when searching for a solution using typical $\chi^{2}$ minimisations, the amount of excess noise in the measurements, or parameter $\sigma_{J}$, is commonly fixed to an \emph{a priori} selected value (e.g. $\sigma_{0}$), which corresponds to setting the prior probability of this parameter equal to a delta-function as $\pi(\sigma_{J}) = \delta(\sigma_{J} - \sigma_{0})$. This prior assumes that only a value of $\sigma_{J} = \sigma_{0}$ is a possible one and ignores the rather likely possibility that this parameter might not be equal to $\sigma_{0}$. If this prior choice was made in the Bayesian context, any scientifically oriented reader would immediately dismiss this choice as too limiting and would not consider the obtained results trustworthy. However, such \emph{prior limitations}, i.e. setting prior densities to zero in chosen subsets of the parameter space, are commonly used in statistical analyses but should be used with care and only when the physical interpretation of the parameters justifies the limitations.

Similar ``hidden'' prior assumptions are made throughout the literature announcing detections of exoplanets around nearby stars. For instance, finding a solution to some data by assuming that planetary eccentricities are equal to zero represents a similar prior choice. It could be argued that in ``dynamically packed'' systems only eccentricities close to zero are viable by enabling the long-term stabilities of these systems, but the possibility that some non-zero orbital eccentricities could also provide stable systems cannot be ignored because enabling eccentricities as free parameters effectively increases the number of free parameter in the model by $2k$ compared to a model with fixed eccentricities and this can have considerable effects on the significances of the obtained solutions. As fully Bayesian, these hypotheses could be easily tested by treating the model with eccentricities fixed to zero as another model in the set of candidate models. In practice, this can be seen as a comparison of different prior densities of the model parameters.

The argument of hidden prior information can in fact be taken even further. Any non-Bayesian analyses correspond to Bayesian ones with the assumption that the prior probability densities of model parameters are actually ``flat'', i.e. uniform distributions in the parameter space. This is a typical uninformative choice for a prior density and is applied widely in the statistical literature to various parameter estimation problems. However, it still \emph{is} a subjective choice and it \emph{does} have an effect on the obtained results. For instance, a typical non-linear transformation of the parameter vector from $\theta$ to $\theta'$ reveals the consequences of this prior choice. If the prior for parameter $\theta$ is a uniform density, that for $\theta'$ has to be chosen by using the Jacobian of the corresponding transformation to receive the same results. If the prior of $\theta'$ is chosen uniform as well, the obtained analysis results would not be the same for these two parameterisations but would depend heavily on the chosen transformation, which emphasises the fact that prior densities are important entities that cannot be ignored when performing statistical analyses. This in turn implies that the only logically consistent framework of statistics is the Bayesian one. While these issues are discussed widely in the statistical literature, we refer to the excellent book \emph{Statistical Decision Theory and Bayesian Analysis} by J. O. Berger \citep{berger1980}.

Because priors are subjective, we present our results only based on the priors we choose. In practice, we adopt the same prior probability densities of model parameters that were used in \citet{tuomi2012}. We also use the same prior model probabilities as \citet{tuomi2012} to take into account the fact that we believe finding $k+1$ planets in any given system is less likely than finding $k$ of them. Because posterior samplings can be used to obtain estimates of the parameter probability densities that enables the computation of any point- and uncertainty estimates, we report all the results using the maximum \emph{a posteriori} (MAP) estimates and the corresponding 99\% Bayesian credibility sets \citep[BCSs; as defined in e.g.][]{tuomi2009b}.

\subsection{Periodograms}\label{sec:periodograms}

Because it is useful to calculate the periodograms of time series, especially when searching for periodic signals, we discuss the resulting power spectra as well for the GJ 581 velocities. Identifying the strongest powers in the periodogram, such as the famous Lomb-Scargle periodogram \citep{lomb1976,scargle1982}, helps choosing the initial states of posterior samplings by pointing towards the regions in the period space where periodic signals are the most likely to be found. The ability to use information from periodogram analyses can help decreasing the burn-in period, i.e. the length of the Markov chain before it has converged to the global solution, of MCMC samplings considerably. However, because periodogram analyses are commonly used to assess the number $k$ in practice, we first discuss the caveats of such assessments.

As noted by \citet{tuomi2012}, periodograms and related analysis methods rely on the assumption that the model residuals have been calculated correctly. For instance, when searching for a $k+1$th signal in a noisy time series, the residuals are commonly calculated by assuming that there are in fact only $k$ periodicities in the data. This obvious contradiction of the assumption and the statistical test performed to search for another signal is not only made throughout the literature when searching for planets using the radial velocity technique, but its effects on the detectability of periodic signals have not received as much attention in the same literature as they should. A clear example is the case of HD 10180 that was reported to host six, possibly seven, planets orbiting it by \citet{lovis2011}. They used power spectra in their attempt of assessing the number $k$ for this star and fell a victim of the above contradiction of the statistical test and the assumption this test is based on. Contrary, the Bayesian analysis of the same exact data performed by \citet{tuomi2012} concluded that the preferred estimate for $k$ is actually as high as nine. This example only demonstrates the weakness of the periodogram-based methods in assessing $k$ but can also be seen as an obvious consequence of the assumption: if it is assumed that there are $k$ signals in a data set, the statistical tests performed based on this assumption will certainly be biased in favour of this assumption being true and prevent the detection of a $k+1$th signal unless it is a strong one and clearly present in the data.

In an attempt to overcome the above shortcomings of periodogram analyses, \citet{anglada2012b} proposed a generalised version of the periodogram defined by \citet{cumming2004} for the purpose of detecting multiple periodicities in time series. This \emph{recursive periodogram} \citep{anglada2012b} can be used to obtain the power spectrum for model residuals by simultaneously adjusting the parameters of the previously detected $k$ signals for each test frequency of the $k+1$th one. Clearly this method is not prone to such a severe contradiction as the traditional periodograms but it is still not free of its own conseptual problems such as the fact that one has to move sequentially by increasing $k$ one step at a time when searching for several periodicities. While certainly an improvement to the classical periodograms, the consequences of hidden prior information cannot be addressed using this modified periodogram either.

We report the results of standard Lomb-Scargle periodogram analyses together with those from the Bayesian analyses. We also report the resulting false alarm probabilities (FAPs) of the signals we detect in the GJ 581 velocities obtained using the HARPS spectrograph.

\section{Sequential analysis of HARPS velocities}

The HARPS velocities provide an interesting opportunity to compare the sensitivity of the Bayesian analysis methods to the traditional periodogram ones used throughout the literature. For this reason, we construct a timeline for GJ 581 where we analyse the available HARPS data after each yearly observing period. We also analyse the data sets of \citet{bonfils2005}, \citet{udry2007}, and \citet{mayor2009} and compare the best estimate of $k$ with an estimate that \emph{could} have been received had Bayesian tools been applied to these velocities since 2005.

We show the number of HARPS velocities after each observing period and at the time when the three studies describing detections of the four planets orbiting GJ 581 and reporting the data were published \citep[these three studies are:][]{bonfils2005,udry2007,mayor1995}. The numbers of measurements and data baselines of these subsets of HARPS data are shown in Table \ref{tab:timeline}. In the following subsections we describe briefly the results of our analyses of the subsets of HARPS data in Table \ref{tab:timeline}. We note, that we assign numbers to the signals we detect by calling the shortest periodicity as signal 1, the second shortest as signal 2 and so forth. We also refer to the signals corresponding to the four planet candidates by using the letters b, c, d, and e as in \citet{mayor2009}.

\begin{table}
\center
\caption{Number of signals detected from the subsets of HARPS data as a function of time as reported in the literature ($k_{1}$), using Lomb-Scargle periodograms ($k_{2}$), and using Bayesian tools ($k_{3}$). The corresponding data baselines ($T_{\rm obs}$) and numbers of data ($N$) are also shown. The subsets correspond to the published data sets and subsets of \citet{forveille2011} data set after each full observing period (FOP).\label{tab:timeline}}
\begin{tabular}{lccccc}
\hline \hline
Data subset & $N$ & $T_{\rm obs}$ [days] & $k_{1}$ & $k_{2}$ & $k_{3}$ \\
\hline
 \citet{bonfils2005} & 20 & 440 & 1 & 1 & 1 \\
 First FOP & 25 & 457 & -- & 2 & 2 \\
 Second FOP & 43 & 827 & -- & 2 & 2 \\
 \citet{udry2007} & 50 & 1050 & 3 & 3 & 3 \\
 Third FOP & 76 & 1197 & -- & 4 & 4 \\
 \citet{mayor2009} & 119 & 1570 & 4 & 4 & 4 \\
 Fifth FOP & 131 & 1904 & -- & 4 & 4 \\
 Sixth FOP & 197 & 2312 & -- & 4 & 5 \\
 \citet{forveille2011} & 240 & 2543 & 4 & 4 & 5 \\
\hline \hline
\end{tabular}
\end{table}

\subsection{Bonfils et al. data}

When analysing the \citet{bonfils2005} data set, our posterior samplings could rapidly verify that it contains the clear planetary signature of GJ 581 b. This signal was seen easily with posterior samplings (with $P(k=1) \approx 5.9 \times 10^{10} P(k=0)$) and periodograms (FAP $< 0.0001$) and \citet{bonfils2005} were also very confident about its existence based on their analyses that we assume\footnote{In fact, \citet{bonfils2005} do not mention the statistical methods they used to obtain their solution.} relied on periodograms and $\chi^{2}$ minimisations.

Our samplings of a two-Keplerian model also spotted hints of a second periodic signal between roughly 10-16 days. The two-Keplerian model received a significantly greater posterior probability (by a factor of 640) than the one-Keplerian one, though the samplings could not constrain the period of the second signal very accurately. Also, parameter $K$ of this second signal did not differ statistically from zero, which forced us to conclude that there is only one signal significantly present in the \citet{bonfils2005} data set. Yet, this signal exists with little doubt because taking it into account increases the model probability by a factor of 1.1$\times 10^{11}$ compared to the model with $k=0$ despite the low number of measurements (see Table \ref{tab:timeline}).

\subsection{First full observing period}

The end of the first full observing period provided only five more spectra and corresponding velocities of GJ 581 than the data of \citep{bonfils2005}. However, this changed the situation with respect to the second signal considerably. The two-Keplerian model became 4.6$\times 10^{4}$ times more probable than the one-Keplerian one, and parameter $K$ of the second signal was significantly greater than zero with $K_{2} =$ 3.6 [1.5, 5.7] ms$^{-1}$. Despite the fact that the period of this signal had a density with several modes between roughly 12 and 16 days, it was well constrained from above and below and thus satisfied all the detection criteria. The second periodic signal was also detected by the periodograms and the corresponding FAP of this signal was 0.0075.

\begin{figure}
\center
\includegraphics[angle=-90,width=0.24\textwidth]{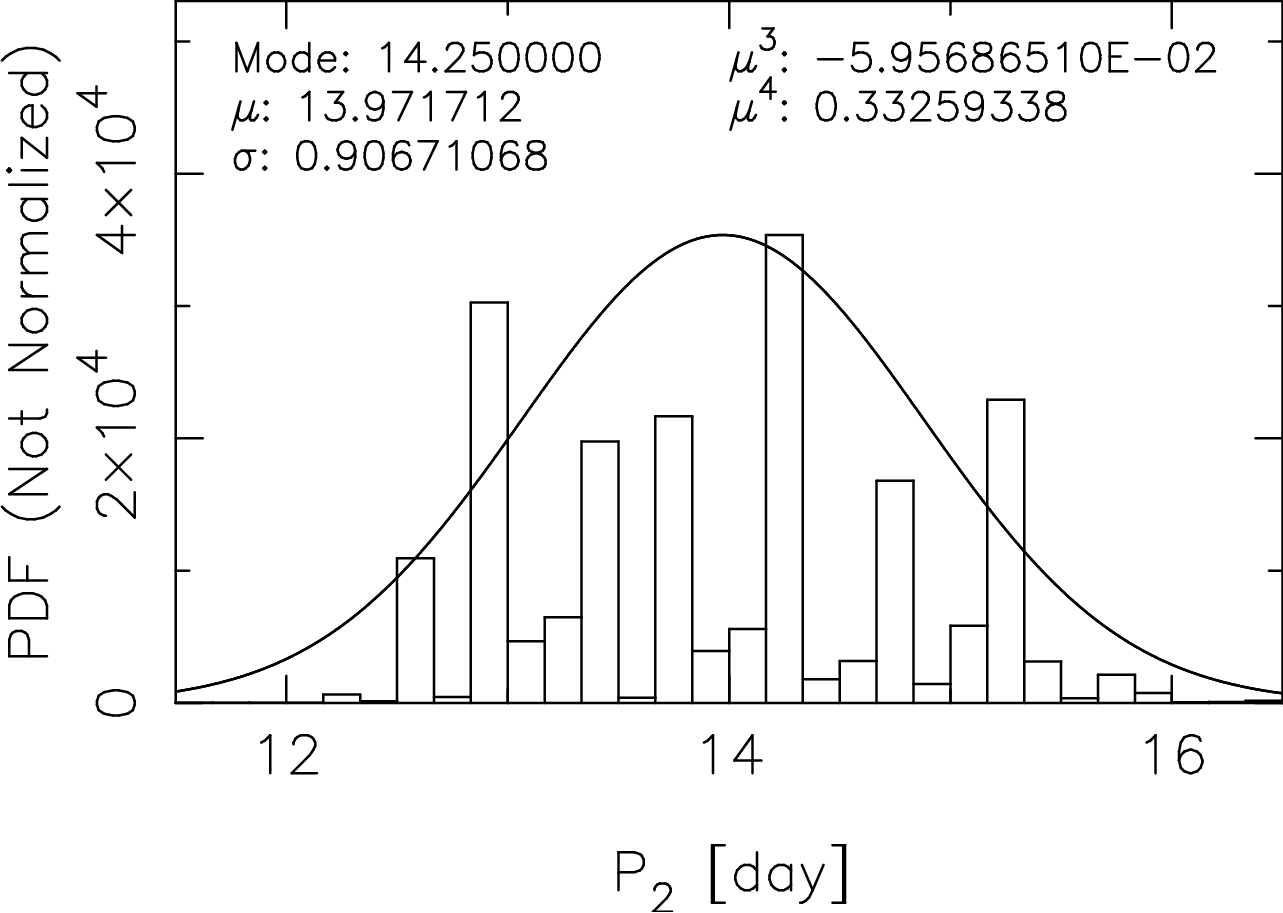}
\includegraphics[angle=-90,width=0.24\textwidth]{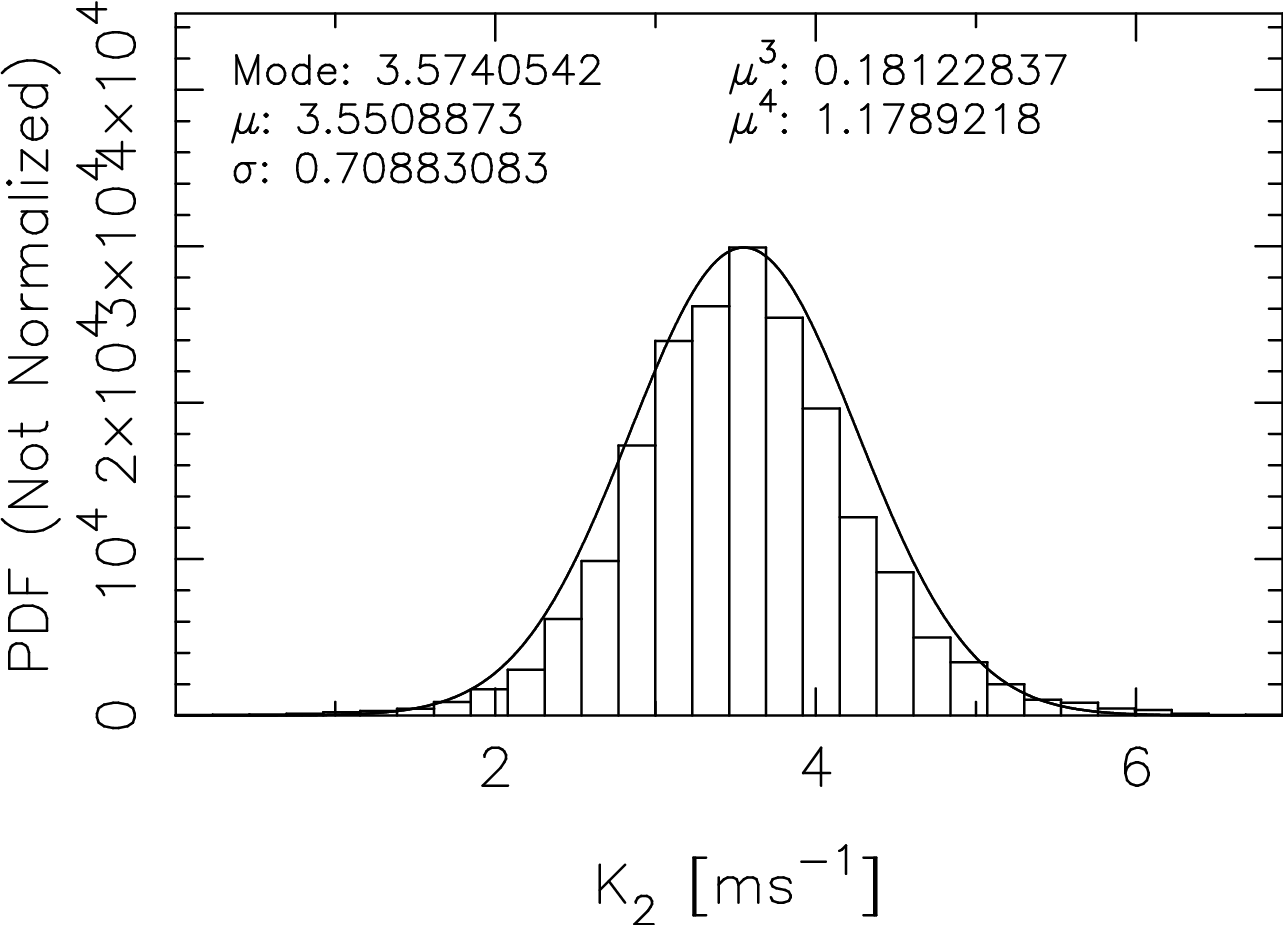}
\caption{Distributions of the velocity amplitude and period of the second signal (corresponding to GJ 581 c) as obtained from the partial HARPS data after the first full observing period. The solid curve shows a Gaussian function with the same mean and variance.}\label{fig:densities_case_b}
\end{figure}

The probability densities of the velocity amplitude and period of the second signal are plotted in Fig. \ref{fig:densities_case_b}. The multimodality of the period is clear from the top panel in Fig. \ref{fig:densities_case_b} but the period is still constrained from above and below and its 99\% BCS is [11.98, 16.27] days. We note that the MAP estimate of the period is 12.9 days.

We sampled a three-Keplerian model as well, but unsurprisingly a third periodic signal did not get constrained by the available 25 velocities.

\subsection{Second full observing period}

At the end of the second full observing period, the amount of velocities had increased to 43, and we could easily sample models with $k=0, ..., 3$. For this data set, the three-Keplerian model implied that there are at least three periodicities in the data but the period of the third signal turned out to have a density with three modes. These modes were found at periods of 42.3, 67, and 84 days. We could not set constraints to the period, and could not be sure whether our Markov chains had indeed converged to this multimodal density -- but assuming they had, the three-Keplerian model would have had a probability of 620 times that of the two-Keplerian one whose parameter space was well sampled and provided two clear periodicities at 5.37 and 12.9 days.

When we sampled the parameter space of a four-Keplerian model, however, we could identify a solution for a fourth period at 3.1 days. This solution, while only suggestive because the period was not constrained from above or below but was only apparent as a global maximum in the otherwise ``flat'' probability density, actually helped constraining the third periodic signal from above and below according to our detection criteria. While this subset of HARPS data could not be used to claim a detection of four periodicities, the results suggested that only a modest increase in the number of available data would be likely to provide detections of four signals. Yet, according to our detection criteria, we could spot only two signals confidently in this data set, namely, those with periods of 5.37 and 12.9 days. These two signals were also detected by the periodogram analyses (Table \ref{tab:timeline}).

\subsection{Udry et al. data}

\citet{udry2007} reported that the \citet{bonfils2005} data already actually suggested there might be an additional companion orbiting GJ 581 with an orbital period of 13 days with a modest significance. With additional 30 data points and improved barycentric corrections, their periodogram analyses and $\chi^2$ minimisations pointed towards two new periodic signals, corresponding to two new planets orbiting GJ 581 with orbital periods of 13 and 84 days.

\begin{figure}
\center
\includegraphics[angle=-90,width=0.24\textwidth]{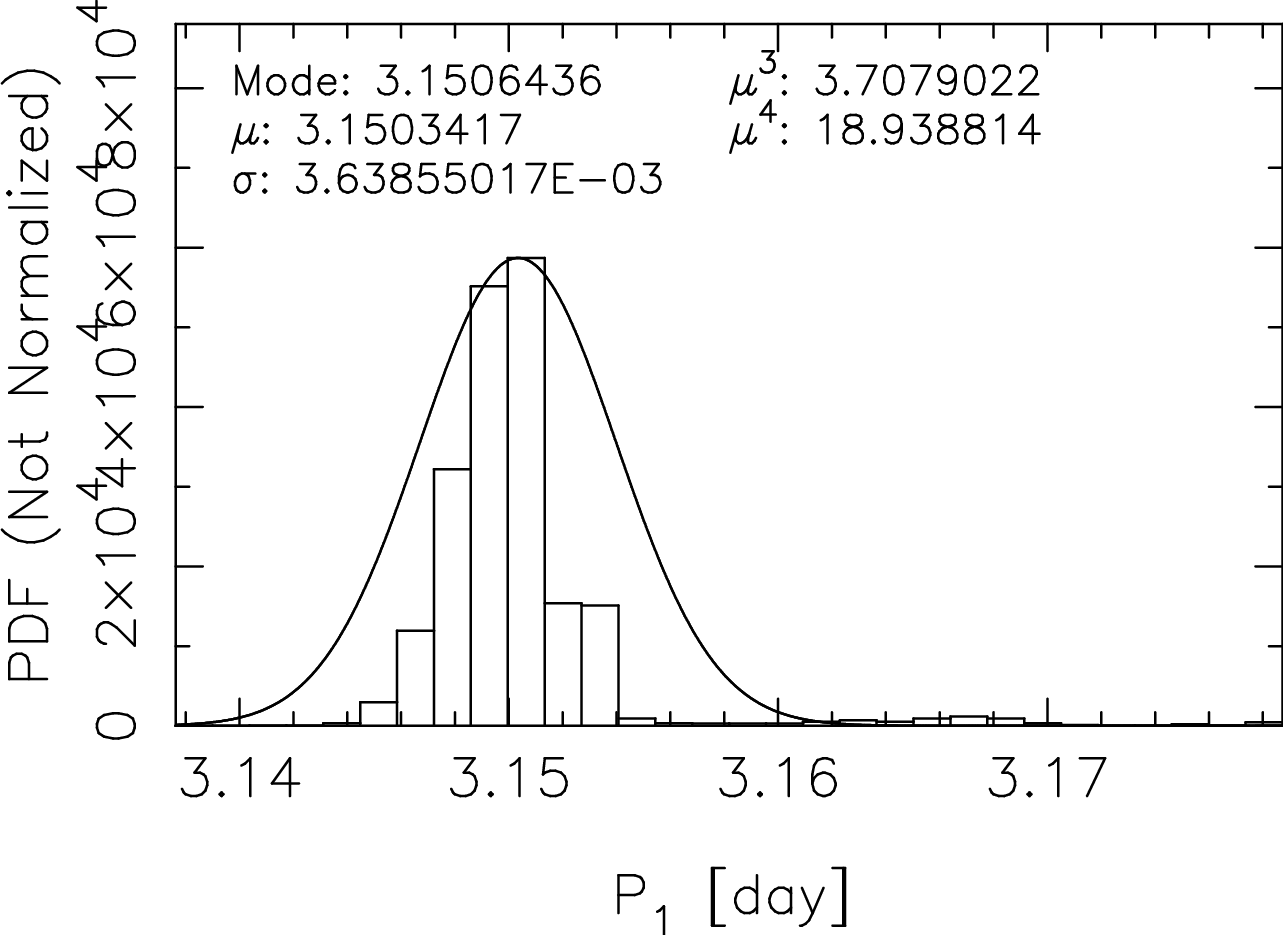}
\includegraphics[angle=-90,width=0.24\textwidth]{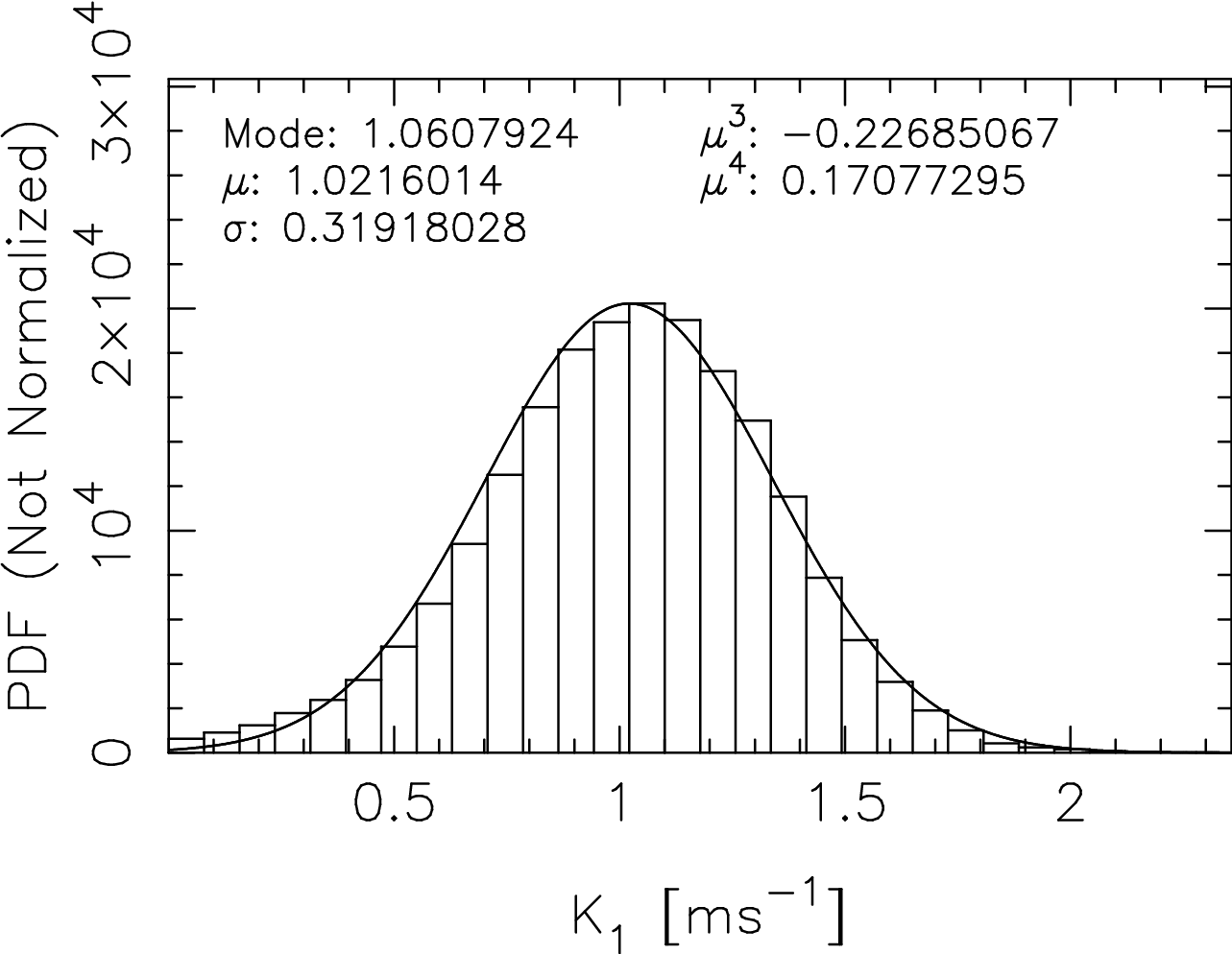}
\caption{As in Fig. \ref{fig:densities_case_b} but for the velocity amplitude and period of the signal with the shortest periodicity (corresponding to GJ 581 e) from the \citet{udry2007} data.}\label{fig:densities_case_d}
\end{figure}

These two signals were indeed significantly present in the \citet{udry2007} data according to our Bayesian analyses. However, we could go further than that and identify a fourth periodicity at 3.14 days corresponding to the fourth planet candidate reported by \citet{mayor2009}. We show the probability densities of velocity amplitude and period of the fourth signal (Fig. \ref{fig:densities_case_d}) to demonstrate that the fourth periodicity was detected as a clear probability maximum but it did not quite satisfy the detection criteria, namely, parameter $K$ of the 3.14 day signal was not significantly greater than zero by having a tail that extended all the way to zero in the amplitude space (Fig. \ref{fig:densities_case_d}, top panel). However, the signals reported by \citet{udry2007} were clearly detected in the data.

The periodogram analyses revealed the existence of three signals in this data set as well. The third most significant signal was found at 82.6 days with a FAP of 0.001 but the interpretation of the origin of this signal corresponding to the reported period of 84 days was less clear than claimed by \citet{udry2007} because the sixth stronges power in the window function was suspiciously close to this period at 87 days making it possible that this periodicity could be due to sampling and not to a genuine signal.

\subsection{Third full observing period}

After the third full observing period with 76 radial velocities, the 3.14 day signal that was close to satisfying the detection criteria in the \citet{udry2007} data set was well constrained and satisfied all the detection criteria. With these velocities, the period of the outer planet candidate received a global maximum at 67 days instead of the periodicity at 84 days in the \citet{udry2007} data set. Essentially, we obtained the same solution that was reported by \citet{mayor2009} after four full observing periods and 43 more velocities.

\begin{figure}
\center
\includegraphics[angle=-90,width=0.24\textwidth]{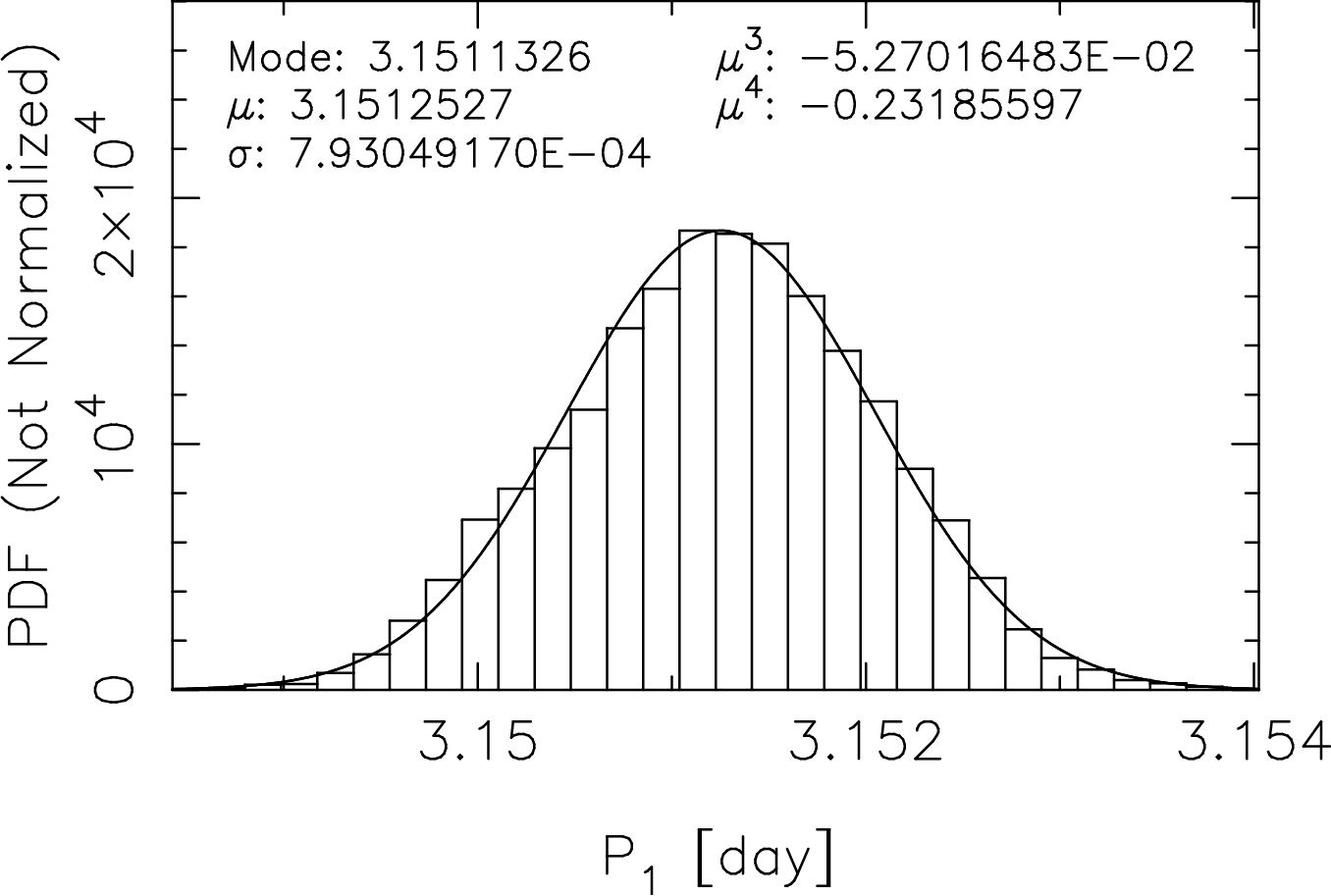}
\includegraphics[angle=-90,width=0.24\textwidth]{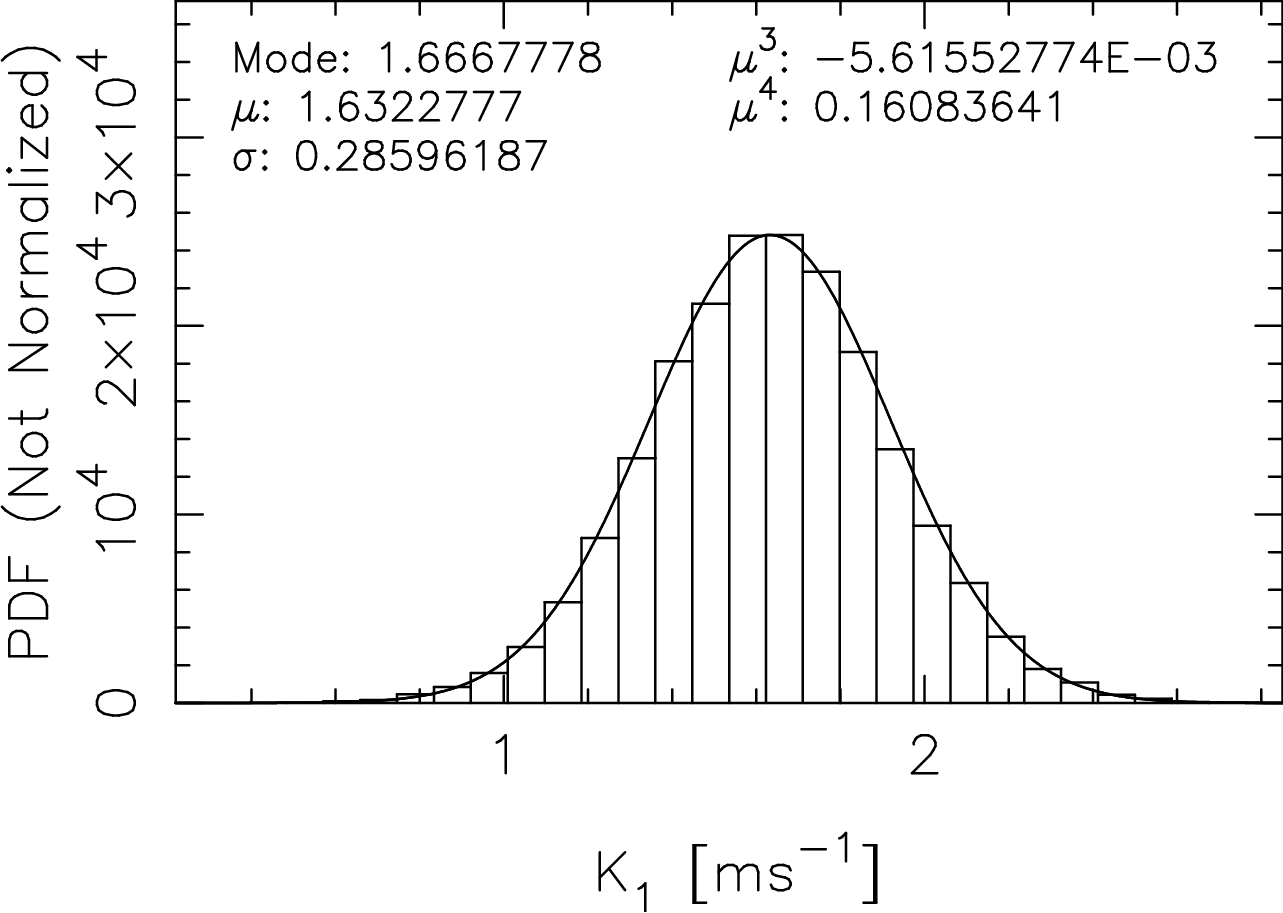}

\includegraphics[angle=-90,width=0.32\textwidth]{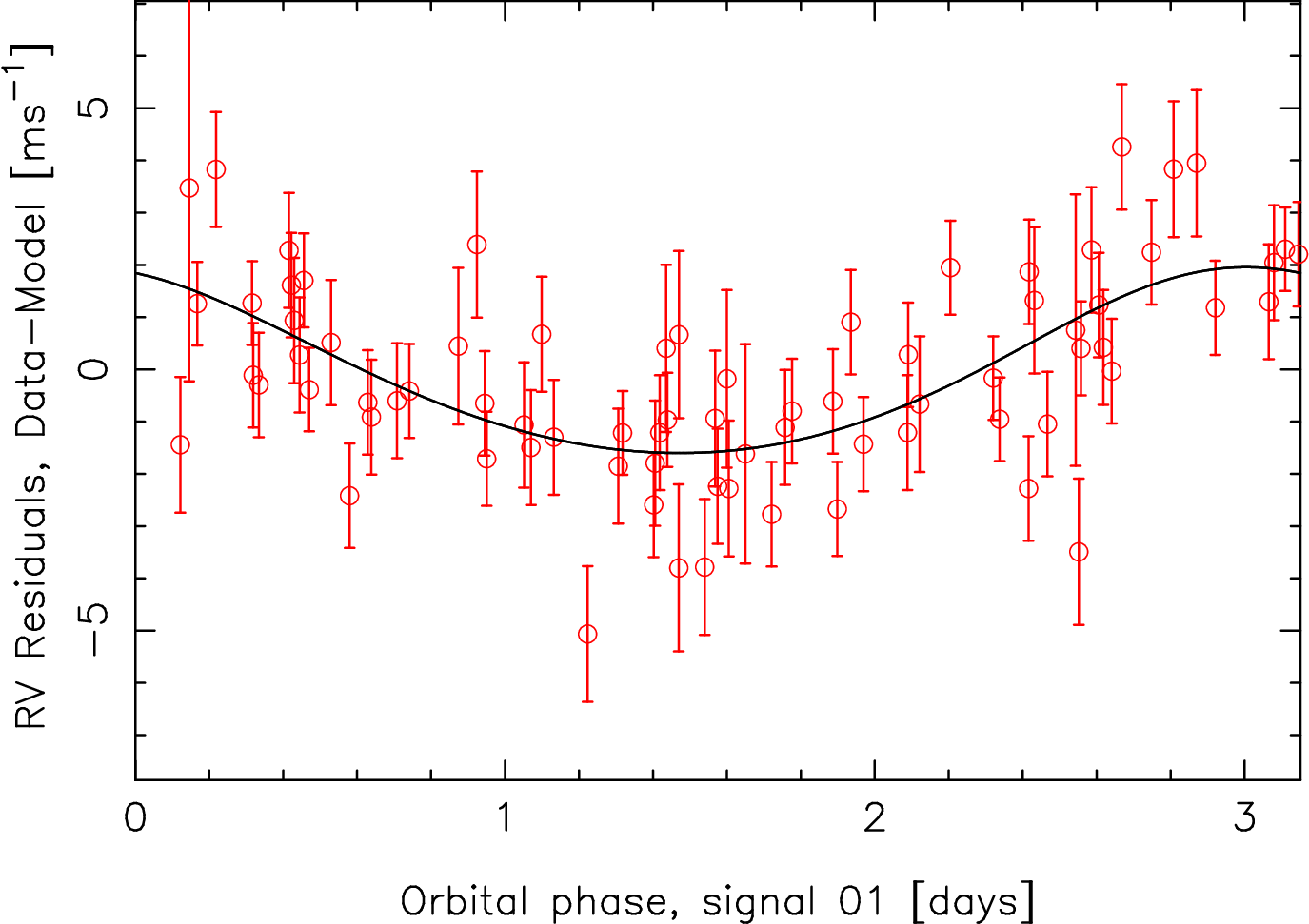}
\caption{Probability densities of parameters $P$ and $K$ of GJ 581 e as in Fig. \ref{fig:densities_case_b} and the phase-folded MAP signal (bottom panel) with the other three signals subtracted.}\label{fig:densities_case_e}
\end{figure}

To demonstrate that the solution indeed satisfied the detection criteria, we show the posterior densities of the period and velocity amplitude of the ``weakest'' 3.14 day signal in Fig. \ref{fig:densities_case_e} (top panels) together with the phase-folded signal (bottom panel). Because the corresponding four-Keplerian solution also had a posterior probability that was 5.4$\times 10^{7}$ times greater than that of the three-Keplerian model, we could conclude that the 76 epochs of HARPS data obtained after three full observing periods were already clearly in favour of a four-Keplerian model. We note that despite the eccentricity prior that penalises high eccentricities, the eccentricity of the outer companion had a MAP estimate of 0.30, yet, its uncertainties were considerable and it was found to be statistically indistinguishable from zero with a 99\% BCS of [0, 0.58]. The four signals were also obtained by the periodogram analyses and the FAP of the fourth signal was 0.001, which indicates that this signal was significantly present in the data.

We sampled the five-Keplerian solution and detected two probability maxima for the fifth signal at 25 and 42 days. However, neither of these was found to correspond to a solution that satisfied the detection criteria of another signal -- the corresponding periods did not get well constrained and the velocity amplitudes received values were not found distinguishable from zero.

\subsection{Mayor et al. data}

\citet{mayor2009} reported the detection of the fourth and innermost planet candidate orbiting GJ 581 with an orbital period of 3.14 days after having observed 119 spectra of the star. This new data set included velocities based on all the existing spectra after improved spectral reductions and corrections accounting for e.g. a changing internal pressure of the Th-Ar calibration lamps caused by aging. \citet{mayor2009} stated that they discovered a fourth periodic signal in the velocities using ``non-linear minimisations with genetic algorithms'' and periodogram analysis.\footnote{\citet{mayor2009} did not actually specify which measure of model goodness they minimised, which makes it impossible to replicate their results. However, the term ``minimisations'' that they used suggests that the chosen measure was $\chi^{2}$ statistics which corresponds to a delta-function prior for the jitter.}

We could obtain the same four-Keplerian solution easily using the \citet{mayor2009} velocities. However, we also tested whether we could spot a fifth signal in this set of 119 radial velocities. Indeed, sampling the parameter space of the five-Keplerian model identified three probability maxima for a fifth periodicity in the data. These maxima were found at periods of 17, 34, and $\sim$400 days but the corresponding Markov chains did not actually converge to a posterior density in the sense that their means and variances did not show converge as a function of the chain length. While this might be a result of insufficient chain length, we drew samples of up to few $10^{8}$ members from the posterior density, which should be a sufficient length given the acceptance rate of roughly 10\% throughout the samplings \citep{tuomi2012}. Also, for the local maxima at 17 and 34 days, the velocity amplitude parameter was not found statistically different from zero, which violates the detection criteria. Similarly, the global maximum of the fifth periodicity at $\sim$400 days (likely corresponding to the periodicity reported by \citet{vogt2010} at 433 days) was found to have an amplitude consistent with zero despite the fact that its MAP estimate was found to be as high as 1.7 ms$^{-1}$. The signal corresponding to this global maximum was also found to have an extremely poor phase-coverage due to annual gaps in the HARPS velocities.

\subsection{Fifth full observing period}

The 131 velocities obtained after the fifth full observing period did not change the solution much with respect to the one obtained using the data set of \citet{mayor2009}. The four signals reported by \citet{mayor2009} were very clearly present in the data and we moved on to sample the parameter space of a five-Keplerian model. This time, we spotted four different probability maxima for the period of the fifth signal at 32, 42, 190, and 420 days. However, none of these maxima corresponded to solutions satisfying the detection criteria by having either posterior probabilities below the detection threshold of by having velocity amplitudes statistically indistinguishable from zero. Therefore, we could not claim that the data supported the existence of more than four signals according to our detection criteria.

Curiously, the 420 day probability maximum corresponds to the signal reported by \citet{vogt2010} whereas the 32 day one corresponds to the signal reported by \citet{vogt2012}. These two probability maxima, despite the fact that they do not satisfy the detection criteria, are therefore existing features of the parameter density of the five-Keplerian model. Based on these 131 velocities, however, it is impossible to state whether these probability maxima correspond to actual signals in the data or false positives in the analyses of \citet{vogt2010} and \citet{vogt2012}.

\subsection{Sixth full observing period}

After the sixth full observing period, the total amount of HARPS velocities for GJ 581 had increased to 197. These data further improved the constrains of the orbits of the four planets detected by \citet{mayor2009} and increased their significances w.r.t. the 131 velocities that had been available a year earlier after the fifth full observing period. When sampling the five-Keplerian model, we could identify an interesting solution for the fifth signal.

\begin{figure}
\center
\includegraphics[angle=-90,width=0.32\textwidth]{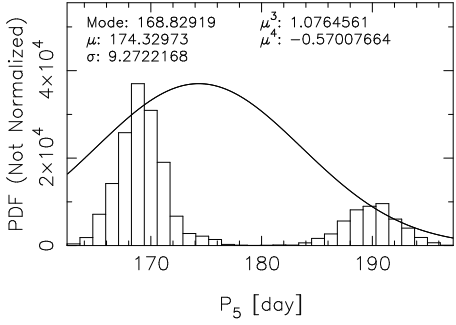}
\caption{Probability density of the fifth periodicity in the HARPS data after the sixth full observing period.}\label{fig:6fop_bimodal}
\end{figure}

According to our samplings, there were actually two probability maxima in the five-Keplerian model corresponding to 170 and 190 day periodicities for the fifth signal. These maxima, by being rather close to one another in the period space, enabled drawing a sample from the posterior density and our Markov chains were found to ``bounce'' between the two maxima yielding a sample from this bimodal solution. This solution satisfied all the detection criteria by having a well constrained period, velocity amplitude that was significantly greater than zero, and a model probability that was 4.7$\times 10^{3}$ times greater than the probability of the four-Keplerian model, which indicates that the favoured value of $k$ is 5 for this data set. While the bimodality is a rather unsatisfactory feature of a probability density -- especially when the goal is to quantify the properties of signals corresponding to candidate planets -- it is by no means surprising considering that we are searching for periodic signals in the first place. Similar bimodal densities have been observed before when dealing with radial velocities \citep[e.g.][]{tuomi2012}. We show the posterior density of the period of the fifth signal in Fig. \ref{fig:6fop_bimodal}. We could not find any other significant periodicities in this data. Also, we could not detect the fifth signal using the periodogram analyses of the data.

\subsection{Forveille et al. data}\label{sec:forveille_data}

\citet{forveille2011} reported that they could spot four and only four periodic signals in their set of 240 HARPS velocities. This was also a simple task for our periodogram analyses as well as for the Bayesian posterior samplings and we continued to sample the parameter space of the five-Keplerian model. These samplings indicated that the 190 day periodicity observed in the data after the sixth observing period gained greater significance while the 170 day period ceased to correspond to a probability maximum. The probability of the five-Keplerian model was found to be 5.7$\times 10^{4}$ times more probable than the four-Keplerian one, which shows that the significance of the fifth periodic signal at 190 days received roughly ten times greater significance than it did for the data after the sixth full observing period. This periodicity was also well constrained and satisfied all our detection criteria. The periodogram analyses could not be used to detect a fifth periodogram in the data confidently.

\begin{figure}
\center
\includegraphics[angle=-90,width=0.24\textwidth]{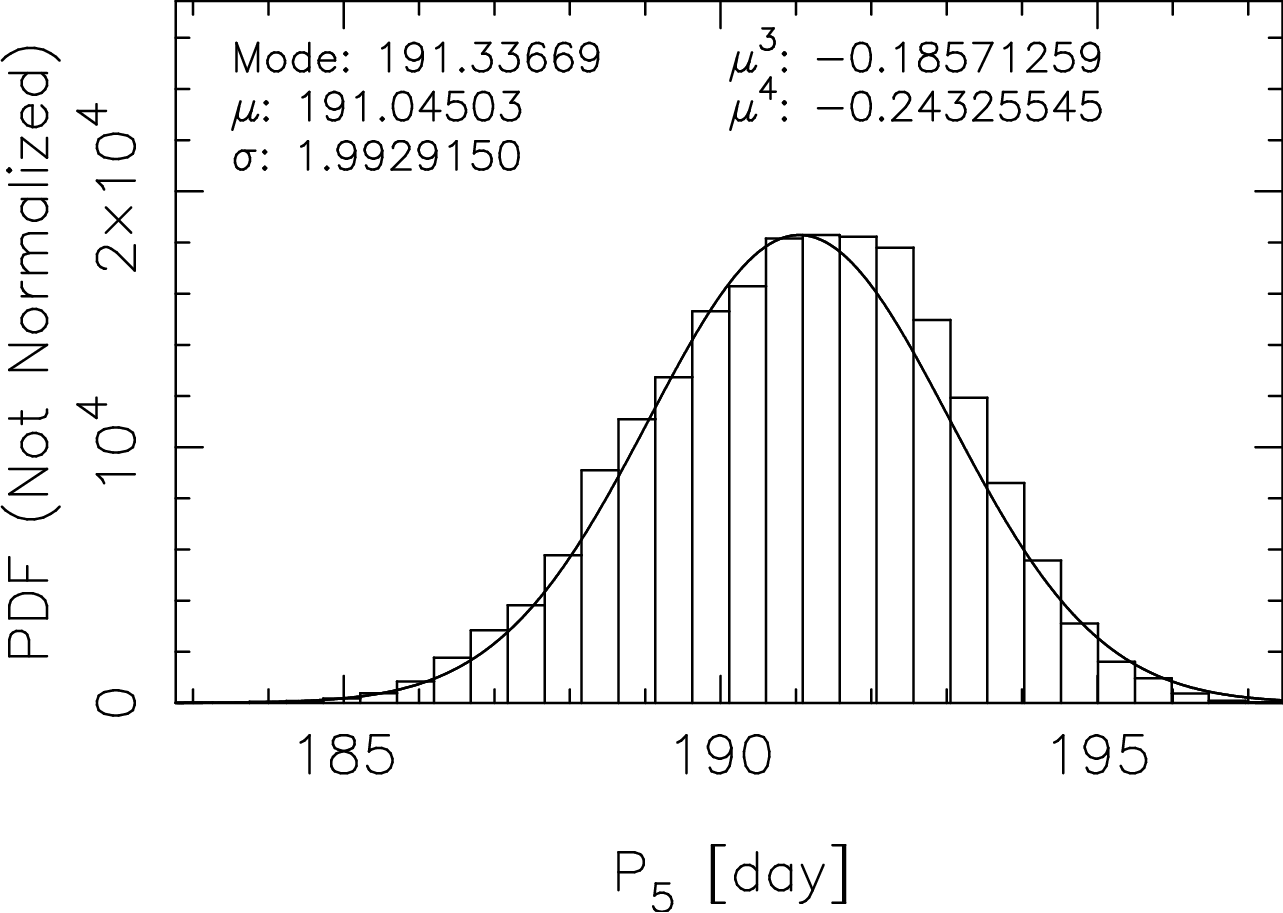}
\includegraphics[angle=-90,width=0.24\textwidth]{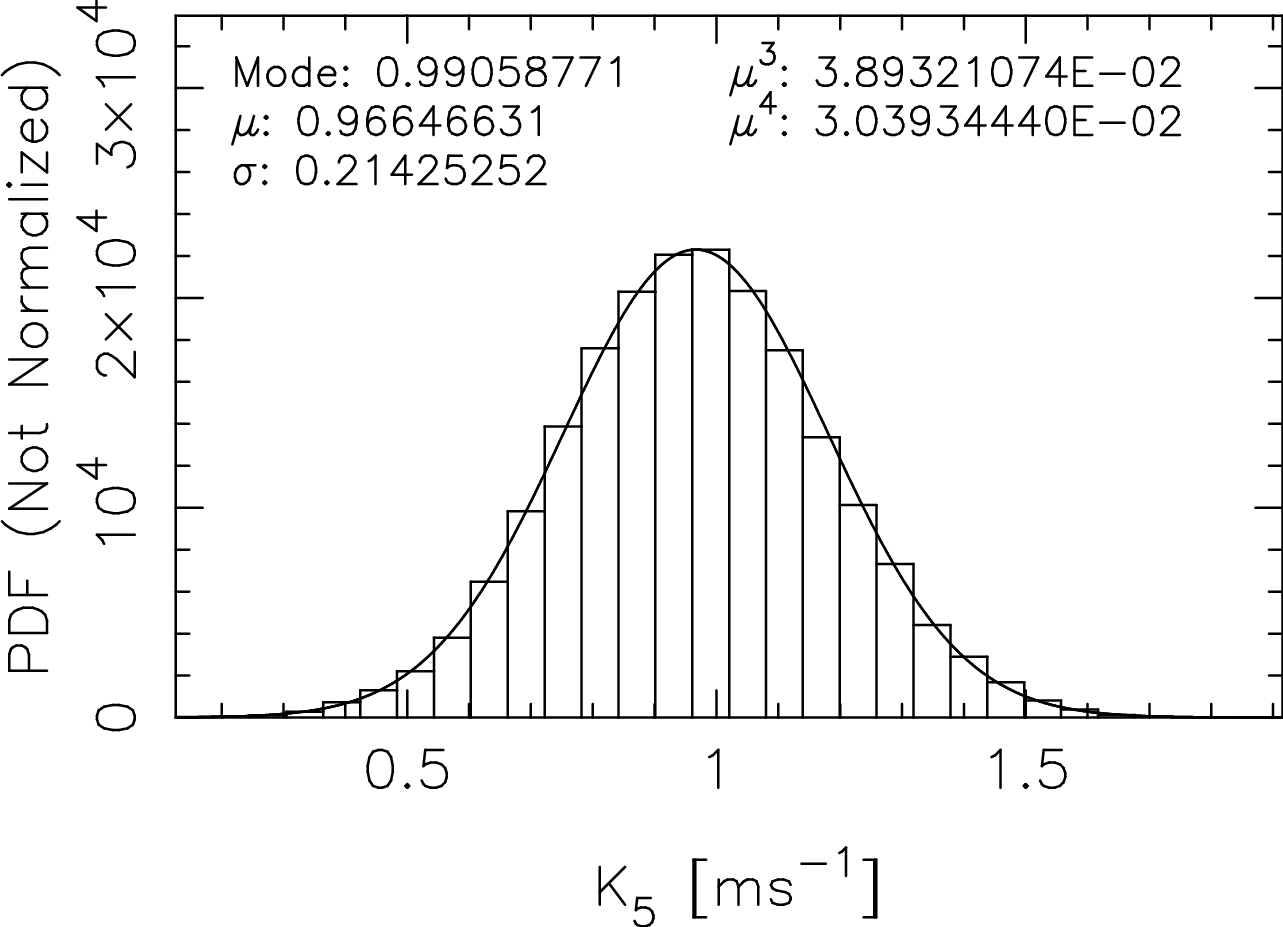}

\includegraphics[angle=-90,width=0.32\textwidth]{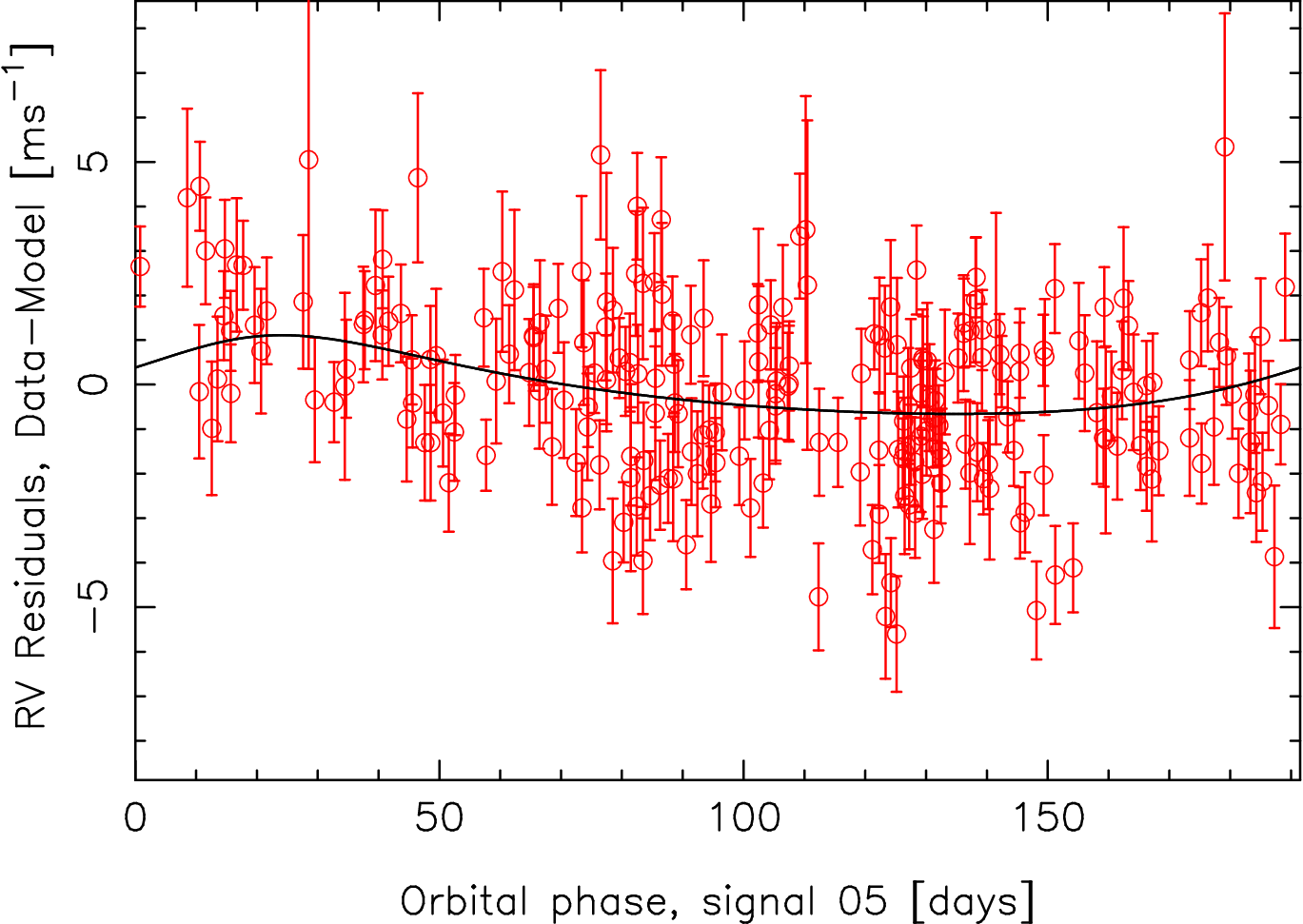}
\caption{As in Fig. \ref{fig:densities_case_e} but for the fifth signal in the \citet{forveille2011} data.}\label{fig:forv_fifth}
\end{figure}

However, as also reported by \citet{vogt2012}, we also spotted another probability maximum corresponding to a periodicity at 32 days. Treating the 32 and 190 day solutions as separate models, we calculated their respective probabilities. These probabilities were found to be 7\% and 93\%, respectively, which indicates that the global solution of the five-Keplerian model corresponds to a fifth periodicity at 190 days. Because of the existence of the probability maximum at 32 days, we performed global samplings of a six-Keplerian model but failed to find a signal satisfying the detection criteria. We did, however, spot a probability maximum for the period of the sixth signal at 32 days, but it could not be constrained from above or below, which violates one of our detection criteria. We show the phase-folded fifth signal in Fig. \ref{fig:forv_fifth} (bottom panel). While the phase-folded signal is not visually very impressive, we also plotted the distributions estimating the period and amplitude of this signal in Fig. \ref{fig:forv_fifth} (top panels) which indicate that this signal complies with our detection criteria.

\section{Analysis of combined radial velocities}

\begin{figure}
\center
\includegraphics[angle=-90,width=0.45\textwidth]{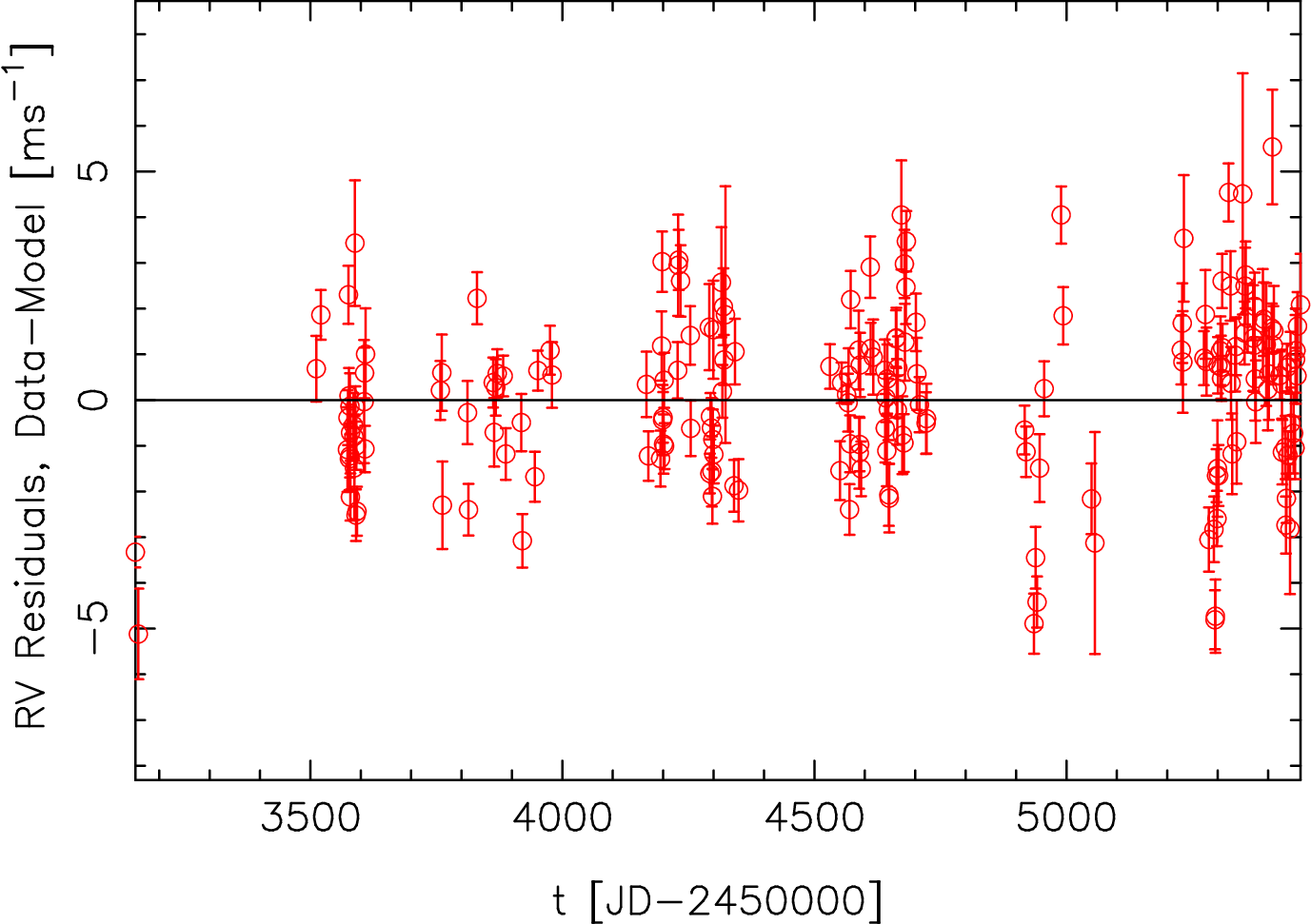}

\includegraphics[angle=-90,width=0.45\textwidth]{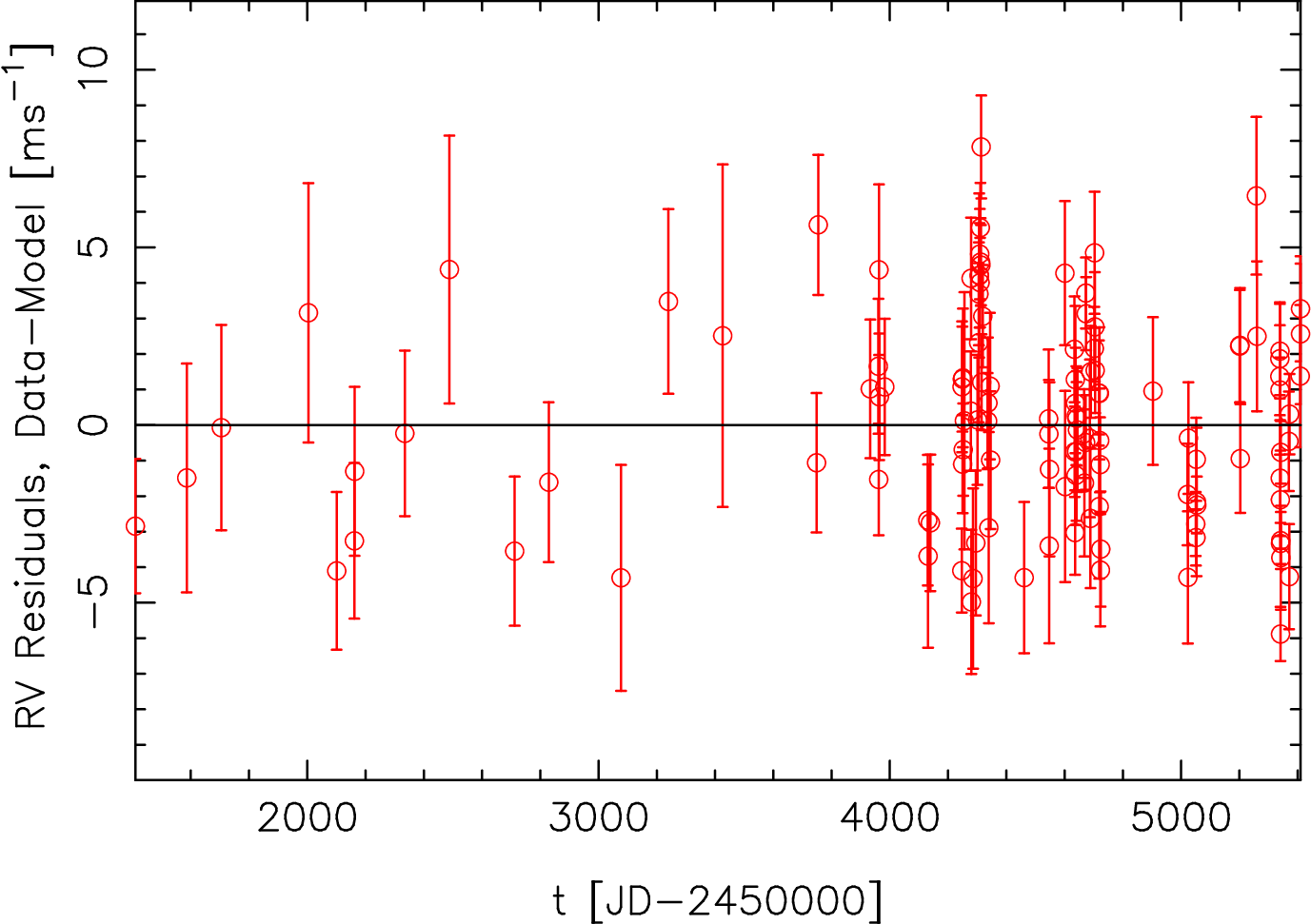}
\caption{HARPS (top) and HIRES (bottom) residuals of the four-Keplerian model.}\label{fig:residuals_4}
\end{figure}

In this section we perform a robust analysis of the combined HARPS and HIRES velocities of GJ 581. The baseline of the available 240 HARPS velocities of \citet{forveille2011} is 2543 days. We plotted the four-Keplerian model residuals of the HARPS velocities in Fig. \ref{fig:residuals_4} (top panel). Clearly, these velocities show the remarkable stability (with an RMS of only 1.98 ms$^{-1}$) of the HARPS spectrograph and indicate that, in addition to obvious yearly gaps, the HARPS data can be expected to provide accurate constraints to possible additional periodic signals in the data.

The residuals of the four-Keplerian model of the HIRES precision velocities of \citet{vogt2010} are also shown in Fig. \ref{fig:residuals_4} (bottom panel). These velocities have an RMS of 2.82 ms$^{-1}$, which implies an impressive precision, but falls short of the precision of the HARPS velocities for GJ 581. The baseline of the HIRES data is 4000 days, which is almost twice that of the HARPS one. Therefore, combining the HIRES and HARPS velocities would certainly provide greater accuracy than either one of them alone by having a greater overall precision and longer baseline. Also, the differences in the time-distribution of these two sets can certainly help to distinguish genuine Doppler periodicities and their aliases from one another and from spurious signals generated by noise and insufficient modelling by providing a better sampling than the individual sets.

\subsection{White noise model}

As we already analysed the HARPS velocities individually in Section \ref{sec:forveille_data}, and because the HIRES velocities have been analysed individually in e.g. \citet{gregory2011}, where it was concluded that they contain two significant periodicities (the two strongest signals at 5.4 and 12.9 days), we move on directly to analyse the combined data set with the pure white noise model.

Using the Gaussian white noise model, the four signals reported by \citet{mayor2009} were indeed found clearly in the combined HARPS and HIRES velocities and our Markov chains converged to the corresponding four-Keplerian solution with ease. All these four signals were well constrained and satisfied our detection criteria.

When sampling the parameter space of the five-Keplerian model, we searched for signals in the period space between 12.9 and 66.7 days and the space beyond 66.7 days up to a cutoff period of $T_{\rm max} = 10T_{\rm obs}$, where $T_{\rm obs}$ is the data baseline, separately and treated them as different models. Effectively, this corresponds to comparing two different prior models, i.e. in this case prior limitations (see Section \ref{sec:bayes}), with period ranges as $P \in [12.9, 66.7]$ and $P \in [66.7, T_{\rm max}]$, where $P$ and $T_{\rm max}$ have the unit of days, with each other. The cutoff period exceeds the data baseline, but we chose such a long upper limit for the parameter space because signals with periods exceeding the data baseline can be detected in radial velocity data \citep{tuomi2009} and this cutoff choice also enables the detection of signals that are only present in the data as trends with no or little curvature. We chose the two other cutoffs of the period space because we expect \emph{a priori} the inner system of GJ 581 to be ``dynamically packed'' in the sense that because of the existence of planets with orbital periods of 3.1, 5.4, and 12.9 days, it is likely that there are no stable orbits for additional planets with orbital periods less than 12.9 days. This leaves two subspaces of the period space where stable orbits could exist, namely, the ones we define above, i.e. orbits between GJ 581 c and d and orbits beyond GJ 581 d.

\begin{figure}
\center
\includegraphics[angle=-90,width=0.45\textwidth]{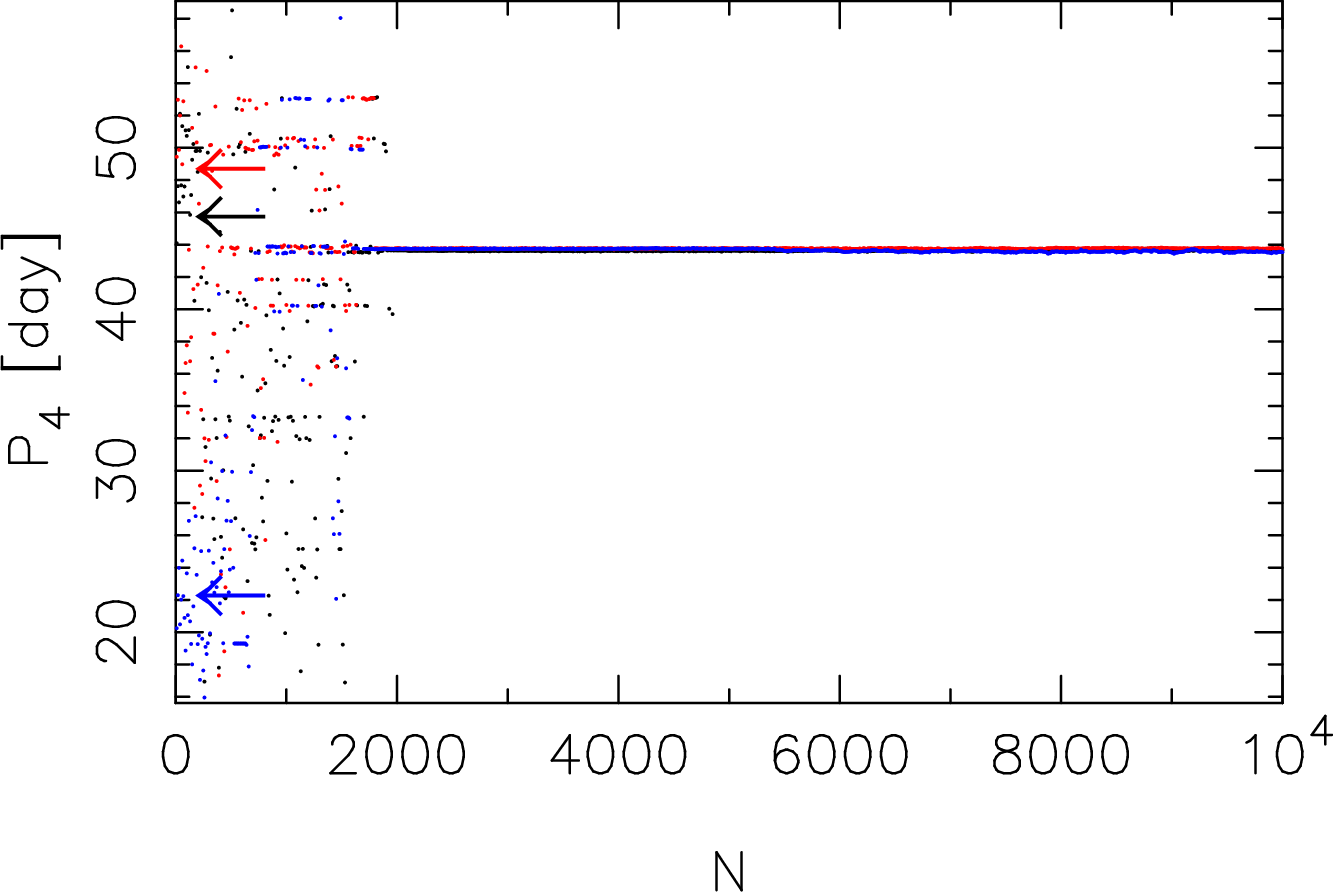}

\includegraphics[angle=-90,width=0.45\textwidth]{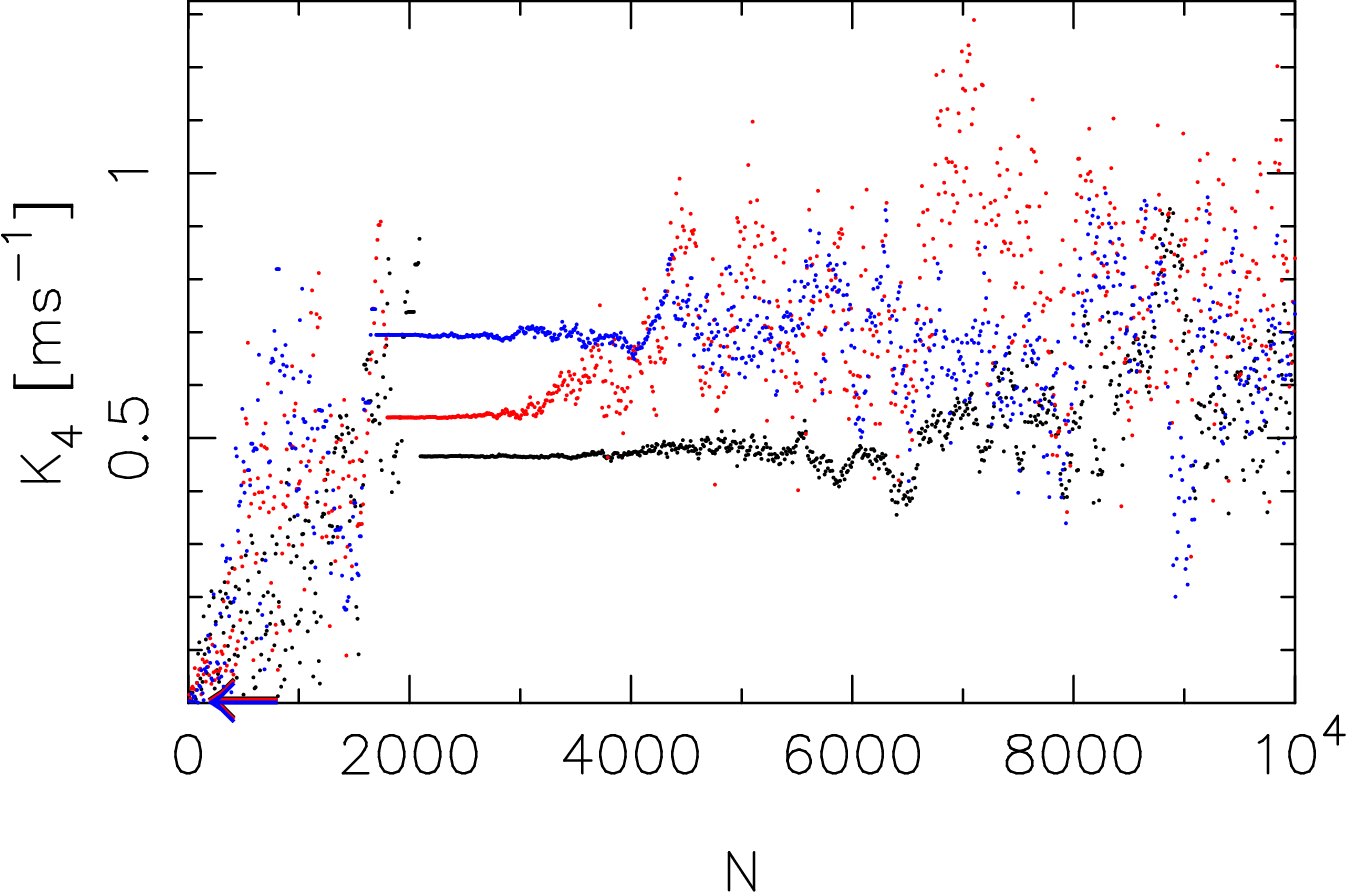}
\caption{Convergence of three Markov chains to a signal with a period of 43.7 days and an amplitude of 0.8 ms$^{-1}$ with randomly selected initial states (arrows) in the period space between 12.9 and 66.7 days given the combined HARPS and HIRES data set and a pure white noise model.}\label{fig:white_fifth_convergence}
\end{figure}

\begin{figure}
\center
\includegraphics[angle=-90,width=0.24\textwidth]{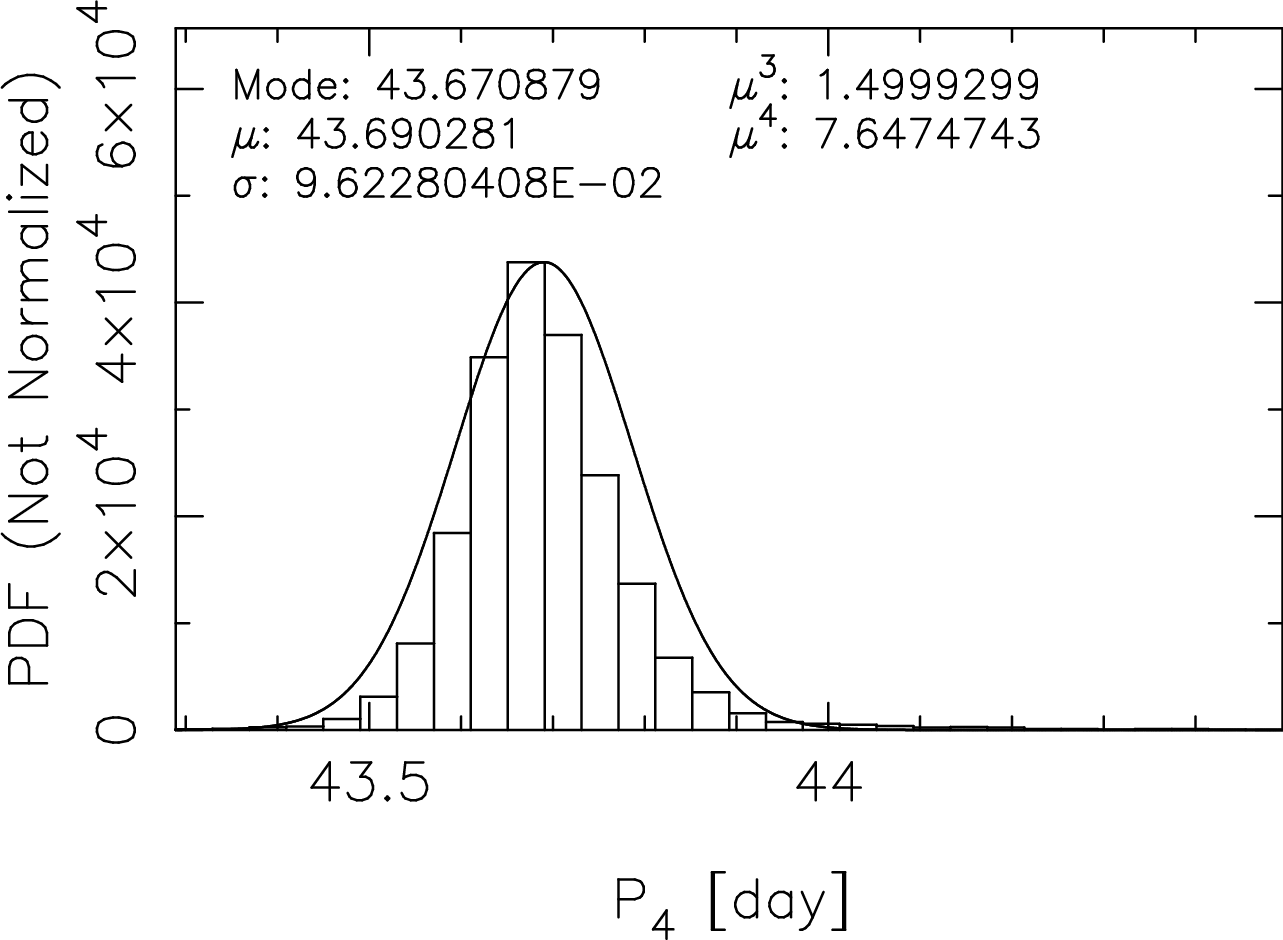}
\includegraphics[angle=-90,width=0.24\textwidth]{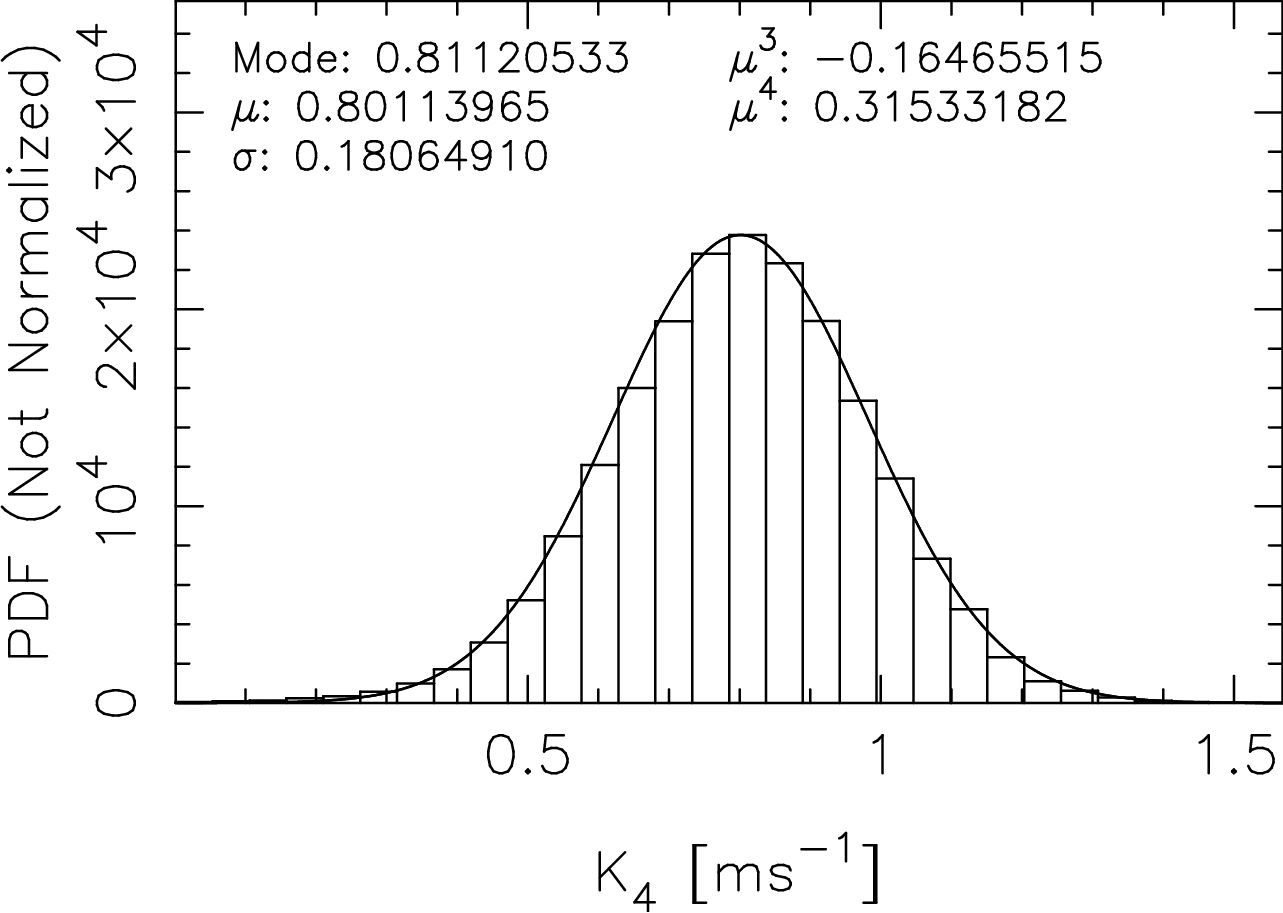}
\caption{Posterior probability densities of the period and amplitude of the fifth most significant (fourth shortest periodicity) signal at a period of 43.7 days given the combined HARPS and HIRES data set and a pure white noise model.}\label{fig:white_fifth_densities}
\end{figure}

Because of the division of the period space into two parts, we could draw samples from both subspaces relatively easily. Samplings of the parameter space of a five-Keplerian model revealed a clear probability maximum for the period of the fifth signal at 43.7 days. We plotted the convergence of the period and amplitude parameters of the fifth signal to this solution in Fig. \ref{fig:white_fifth_convergence} and the posterior probability densities of these two parameters in Fig. \ref{fig:white_fifth_densities}. This signal was significantly present in the data and the five-Keplerian model was found to have a posterior probability of 460 times that of the four-Keplerian one, which exceeds the detection threshold of 150. We note that around the 2000th member of the Markov chains in Fig. \ref{fig:white_fifth_convergence} (bottom panel), the variance of the proposal density has been decreased heavily to increase the convergence rate because the adaptation of the proposal density to the posterior is very low when the chain discovers a narrow probability maximum in the parameter space, as happens in this particular case because the width of the posterior density of the period parameter is much narrower than the corresponding parameter space (see Fig. \ref{fig:white_fifth_convergence}, top panel).

Samplings of both five- and six-Keplerian models revealed an additional probability maximum in the period space between 160-200 days with two maxima at roughly 170 and 190 days, but the posterior samplings did not constrain the amplitude of this signal from below, which violates the detection criteria. Therefore, using the white noise model leads to the conclusion that there are five periodic signals in the combined HARPS and HIRES data. The fact that the 190 day signal observed in the HARPS data alone did not satisfy the detection criteria is unfortunate. While it did appear as a probability maximum given the combined data as well, it could not be constrained as it was for the HARPS data alone. This suggests that this signal is not a genuine Doppler signal of planetary origin but caused by either pure chance in the HARPS velocities or by some (pseudo) periodic source of noise or bias in the HARPS data not accounted for by the pure white noise model.

On the other hand, the nature of the 43.7 day signal in the combined data set is unknown and we cannot rule out its origin as a genuine Doppler signature of planetary origin. It is always possible that a signal cannot be detected independently in any of the available datasets but only in the combined set because of the better phase-coverage, longer baseline, and a greater number of measurements in the combined set. Yet, it could also be caused by systematic differences between the different data sets, such as small instrumental biases that give rise to weak signals in the combined data set when not accounted for. However, we do not find any indication of such biases and in fact the HARPS and HIRES data sets cannot be shown to contradict with one another given the five-Keplerian model based on the Bayesian model inadequacy criterion of \citet{tuomi2011b}. Further, it might be a coincidence that the fifth signal appears at a period that, if of planetary origin, implies a 3:2 commensurability of the corresponding planet with GJ 581 d that has an orbital period of 66.7 days. Such an orbital configuration would likely enable the stability of the system in long term but, while this is compatible with a hypothesis that the 43.7 day signal is caused by a fifth planet candidate in the system, it does not imply the existence of five planets around GJ 581 unless the existence of the fifth signal is supported by future data. Whether this explanation for the 43.7 day signal is valid or not also remains to be investigated with better noise models.

We note that the excess noise levels (estimates of parameter $\sigma_{J}$) of the HARPS and HIRES velocities were 1.50 [1.15, 1.86] and 2.20 [1.63, 2.90] ms$^{-1}$, respectively, given the best model with $k=5$.

\subsection{Red noise model}

We tested the dependence of the observed signals on the selected noise model and the conclusions of \citet{baluev2012} by taking into account red features in the radial velocity noise using a moving average model (MA). This model was constructed as in \citet{tuomi2012c} and \citet{tuomi2012d} but had only one component, making the model a first order MA model with Gaussian white excess noise. Mathematically, this noise model can be written as
\begin{equation}\label{eq:moving_average}
  m_{i} = r_{k}(t_{i}) + \gamma + \epsilon_{i} + \phi \big[ m_{i-1} - r_{k}(t_{i-1}) - \gamma \big] \exp \big[ \alpha (t_{i-1} - t_{i}) \big],
\end{equation}
where each measurement $m_{i}$ at epoch $t_{i}$ is described using the superposition of $k$ Keplerian signals ($r_{r}$); a reference velocity ($\gamma$); Gaussian random variable corresponding to the white noise in the data ($\epsilon_{i}$) with zero mean and variance $\sigma_{i}^{2} + \sigma_{J}^{2}$, where $\sigma_{i}$ is the estimated instrument uncertainty from the spectral reduction pipeline and $\sigma_{J}$ is an unknown excess noise component sometimes referred to as ``stellar jitter''; parameter $\phi$ is a parameter describing the MA magnitude and the term in brackets corresponds to an exponential decrease of this magnitude as a function of time difference of the $i$th and $i-1$th measurements in accordance with the formulation of \citet{baluev2012}. Parameter $\alpha$ describes the timescale of the correlations in the measurement noise.

\begin{figure}
\center
\includegraphics[angle=-90,width=0.24\textwidth]{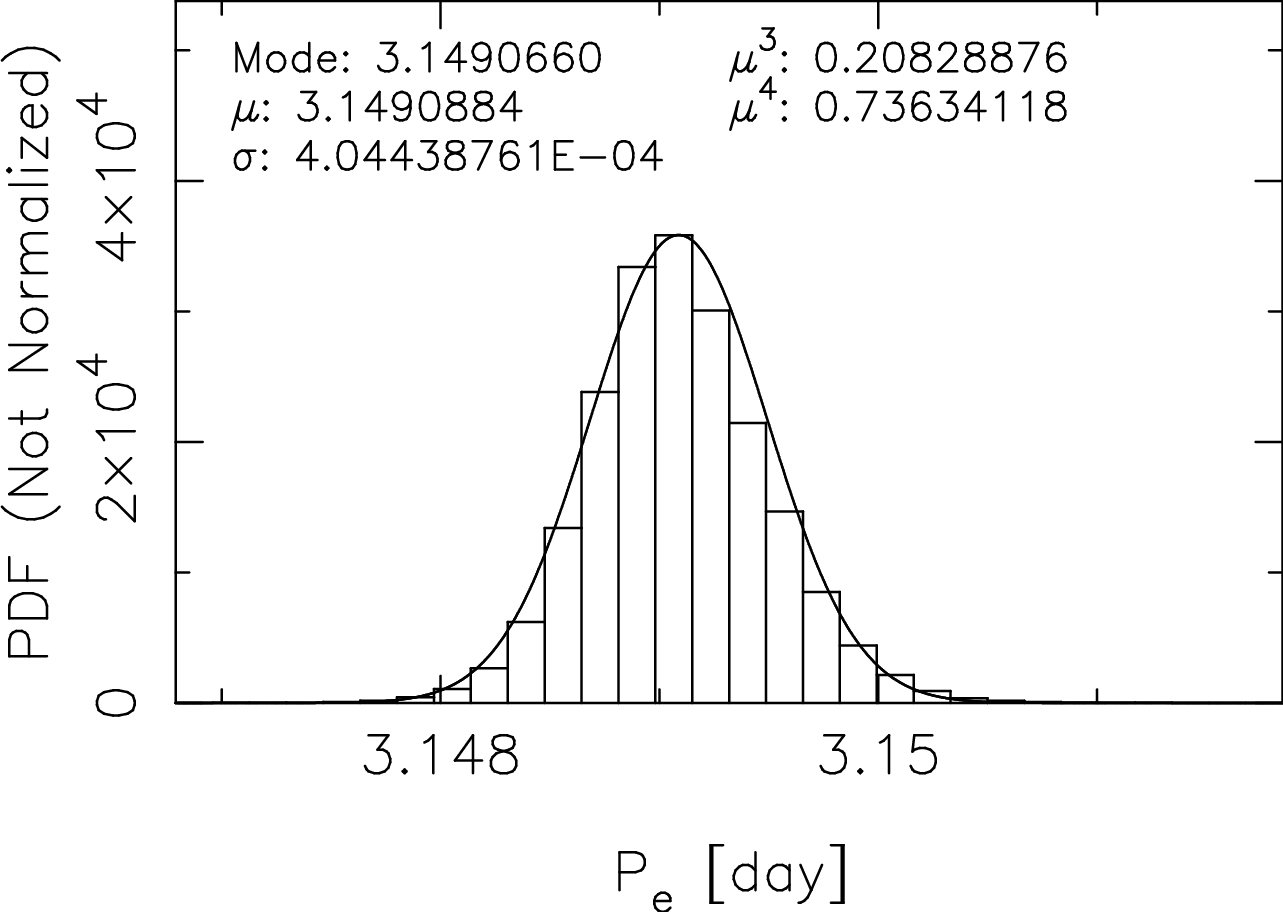}
\includegraphics[angle=-90,width=0.24\textwidth]{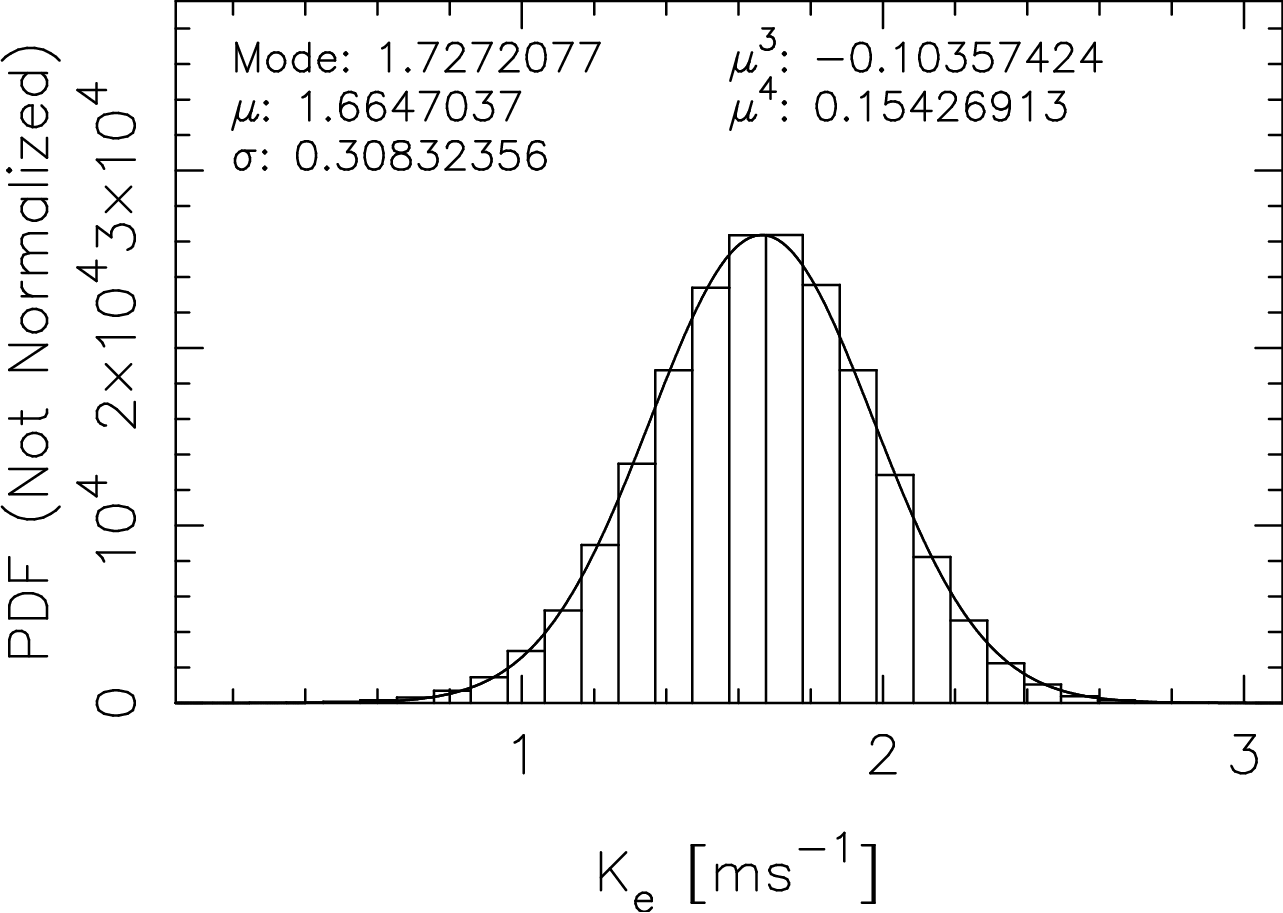}
\caption{As in Fig. \ref{fig:densities_case_b} but for the velocity amplitude and period of GJ 581 e as obtained from the 122 HIRES velocities of \citet{vogt2010} data using the MA model.}\label{fig:densities_HIRES}
\end{figure}

As observed by \citet{baluev2012}, the HIRES data alone actually supports the existence of the planets GJ 581 b, c, and e despite the fact that only b and c could be detected using a pure white noise model and Bayesian analyses of the data \citep{gregory2011}. Using the MA model described above, the samplings of the parameter space of the three-Keplerian model showed beyond reasonable doubt that these three signals are present in the 122 HIRES velocities of \citet{vogt2010} by increasing the posterior probability of the model by a factor of 8.2$\times 10^{5}$ greater than that of the two-Keplerian model that corresponded to the two signals at 5.4 and 12.9 days. The third signal with a period of 3.1 days was constrained well and found consistent with the corresponding signal obtained from the HARPS velocities. We show the posterior densities of the period and velocity amplitude of GJ 581 e in Fig. \ref{fig:densities_HIRES} to demonstrate the significance of this detection.

We could not spot a fourth significant periodic signal in the HIRES data using the posterior samplings and the MA model. However, we performed samplings anyway and spotted three probability maxima for the period of the fourth signal at roughly 70, 300, and 500 days. However, we could not constrain any solutions for the four-Keplerian model and essentially ended up drawing samples from the prior of the period, though the chains were attracted to these three maxima throughout the samplings.

Using the MA model also decreased the estimated amount of white noise in the data. While this white noise was found to be 2.20 [1.63, 2.90] ms$^{-1}$ as estimated with the parameter $\sigma_{J}$ in the pure white noise model, it decreased to 1.81 [1.37, 2.72] ms$^{-1}$ for the MA model.

\begin{figure}
\center
\includegraphics[angle=-90,width=0.24\textwidth]{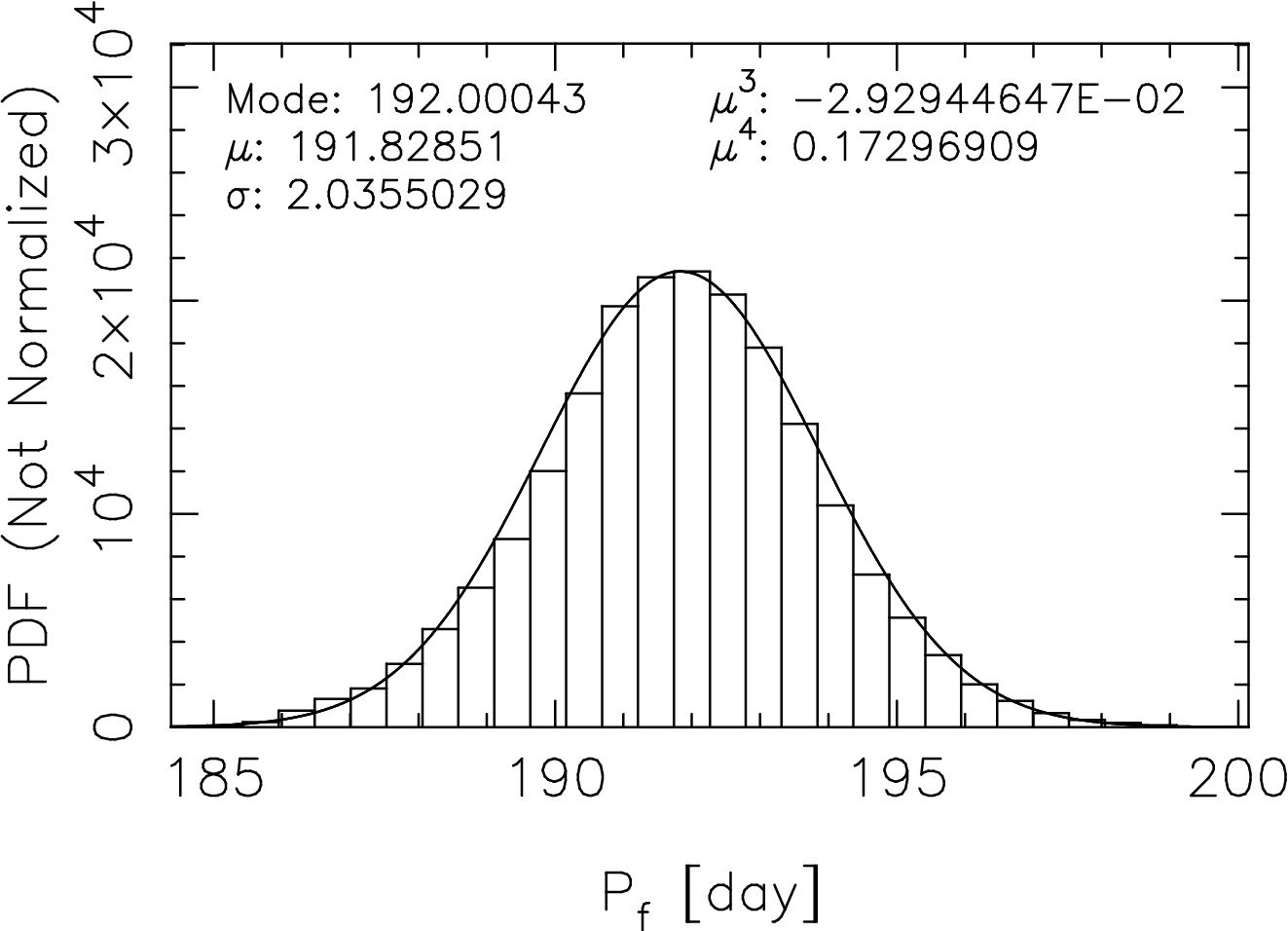}
\includegraphics[angle=-90,width=0.24\textwidth]{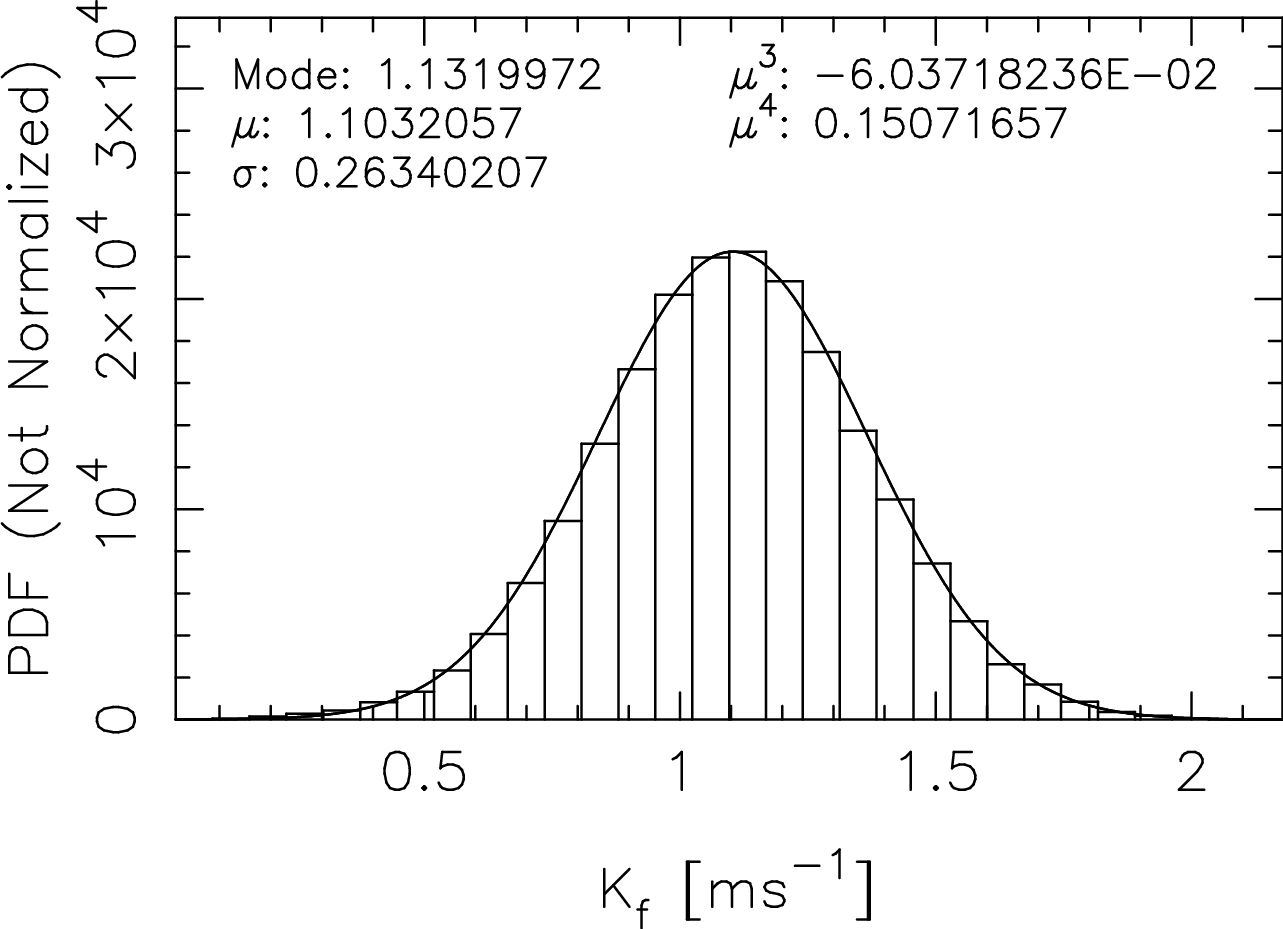}

\includegraphics[angle=-90,width=0.32\textwidth]{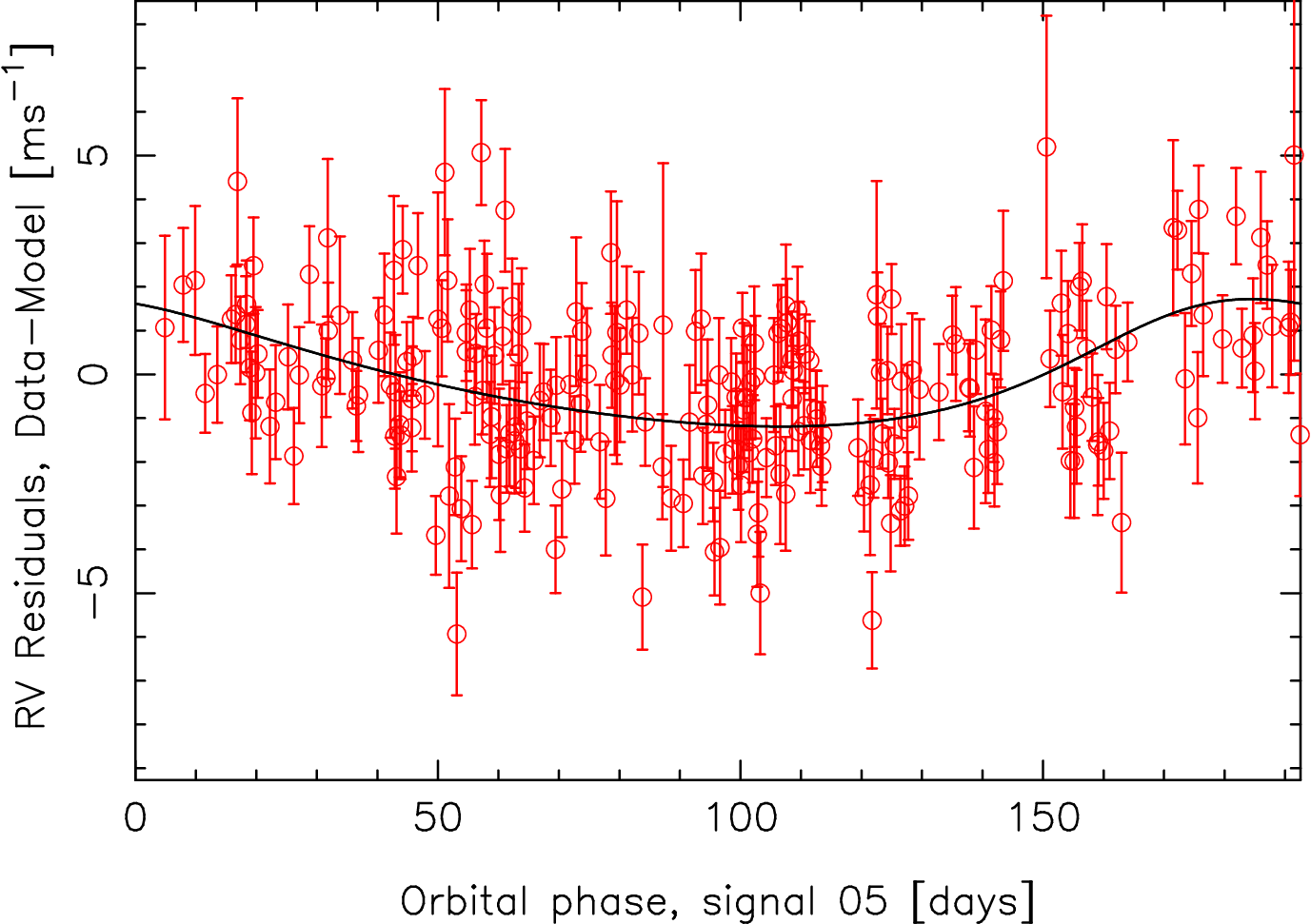}
\caption{As in Fig. \ref{fig:densities_case_e} but for the fifth signal in the \citet{forveille2011} data given the MA model of Eq. (\ref{eq:moving_average}).}\label{fig:MA_fifth_signal}
\end{figure}

The HARPS velocities, however, did appear to favour a model with five periodic signals. We spotted a 190 day periodicity in the HARPS velocities using the MA model of Eq. (\ref{eq:moving_average}). Compared to the four-Keplerian model, the five-Keplerian one had 540 times greater posterior probability and the period and amplitude of the signal at 190 days were well constrained in accordance of our detection criteria. Therefore, the HARPS velocities, as was the case for the analyses with the pure white noise model, supports the existence of five independent sources of periodicity corresponding to the four planets reported in \citet{mayor2009} and one previously unknown signal at roughly 190 day period. We plotted the densities of parameters $P$ and $K$ of this signal in Fig. \ref{fig:MA_fifth_signal} (top panels) together with the corresponding phase-folded signal with the four planetary signatures of GJ 581 b, c, d, and e, and the MA component of the noise removed (Fig. \ref{fig:MA_fifth_signal}, bottom panel). The white noise parameter $\sigma_{J}$ decreased from 1.50 [1.15, 1.86] to 1.19 [0.98, 1.64] ms$^{-1}$ when adding the MA component to the statistical model. This indicates, as was found for the HIRES data, that there are significant correlations in the noise of the HARPS data as well and they have to be accounted for when modelling the velocities.

When analysing the combined HARPS and HIRES data set with the MA model of Eq. (\ref{eq:moving_average}), we did not have any trouble recovering the signals of the four planets orbiting the star. However, as was the case with the white noise model, we failed to identify the existence of the 190 day signal as a significant one when sampling the period space above 67 days with a five-Keplerian model. We did observe a probability maximum at 190 days in the period space but could not constrain the corresponding signal from above and below in the period space and below in the amplitude space in violation of the detection criteria. Also, we did not find a signal at a period of 43.7 days, which indicates that the corresponding periodicity detected using a pure Gaussian white noise model was in fact most likely a spurious signal generated by insufficient noise modelling. This conclusion is supported by the fact that for $k=4$ the MA model was found to have a Bayesian evidence of -797.0 on the logarithmic scale, whereas the corresponding Bayesian evidence of the white noise model was -813.1. Further, the log-Bayesian evidence of the white noise model with $k=5$ was only -806.3, which implies that the MA model with $k=4$ is the best description of the data out of the set of models tested. We show the parameter MAP estimates and their 99\% BCSs of this best model in Table \ref{tab:parameters}.

\begin{table*}
\center
\caption{Four-Keplerian solution of the HARPS and HIRES velocities of GJ 581 together with the parameters quantifying the noise properties. MAP estimates of the parameters and their 99\% BCSs.\label{tab:parameters}}
\begin{tabular}{lcccc}
\hline \hline
Parameter & e & b & c & d \\
\hline
$P$ [days] & 3.14912 [3.14865, 3.14963] & 5.36863 [5.36840, 5.36884] & 12.9169 [12.9104, 12.9235] & 66.63 [66.31, 66.92] \\
$e$ & 0.16 [0, 0.36] & 0.03 [0, 0.05] & 0.03 [0, 0.20] & 0.17 [0, 0.51] \\
$K$ [ms$^{-1}$] & 1.72 [1.36, 2.11] & 12.63 [12.24, 13.05] & 3.21 [2.64, 3.73] & 1.87 [1.12, 2.47] \\
$\omega$ [rad] & 0.6 [0, 2$\pi$] & 0.9 [0, 2$\pi$] & 1.3 [0, 2$\pi$] & 6.2 [0, 2$\pi$] \\
$M_{0}$ [rad] & 5.8 [0, 2$\pi$] & 5.9 [0, 2$\pi$] & 1.9 [0, 2$\pi$] & 0.6 [0, 2$\pi$] \\
$a$ [AU] & 0.0285 [0.0266, 0.0302] & 0.0407 [0.0380, 0.0431] & 0.0731 [0.0684, 0.0773] & 0.218 [0.204, 0.231] \\
$m_{p} \sin i$ [M$_{\oplus}$] & 1.8 [1.3, 2.3] & 15.9 [13.7, 17.9] & 5.4 [4.2, 6.5] & 5.2 [3.3, 7.5] \\
\hline
Parameter & HARPS & HIRES \\
\hline
$\gamma$ [ms$^{-1}$] & 0.11 [-0.46, 0.61] & 0.14 [-0.87, 1.22] \\
$\sigma_{J}$ [ms$^{-1}$] & 1.23 [1.06, 1.72] & 1.62 [1.15, 2.32] \\
$\alpha$ [day$^{-1}$] & 0.057 [0, 0.446] & 0.120 [0, 0.863] \\
$\phi$ & 0.49 [0.26, 0.95] & 0.91 [0.45, 1] \\
\hline \hline
\end{tabular}
\end{table*}

We note that samplings of a six-Keplerian model did not show any evidence in favour of additional periodic signals in the combined data set.

\section{Discussion}

Estimating the number of planetary companions orbiting GJ 581, or equivalently, estimating the parameter $k$ favoured by the data for this star, has been attempted by several authors over the past few years \citep[e.g.][]{bonfils2005,udry2007,mayor2009,vogt2010,tuomi2011,gregory2011,vogt2012,baluev2012}. Improvements in statistical analysis techniques have proven efficient means of improving this estimate, i.e. especially showing that some signals have actually been false positives, but the most significant developments have relied on the availability of new high-precision data. We have assessed the number of significant periodic signals in the combined HARPS and HIRES data of GJ 581 and conclude that there is evidence in favour of four signals, corresponding to the four planets orbiting the star observed by \citet{mayor2009}.

By re-analysing sequentially the HARPS data after each observing period, i.e. each full year, we have shown that the Bayesian statistical techniques were at least as sensitive in detecting periodic signals in the data as the more commonly applied periodogram-based analysis tools (Table \ref{tab:timeline}). Moreover, we did not find any significant periodicities in these HARPS data subsets that did not show as significant signals when the amount of data was increased. This implies that the Bayesian detection criteria \citep{tuomi2012} are not sensitive to false positives arising from noise by chance in practice and any periodic signals that satisfy these criteria are likely genuine signals present in the data.

However, insufficient modelling and possible biases in any single data set can produce false positives that might be interpreted as planetary signals and could contaminate the statistics of exoplanet properties. For instance, the HARPS data was found to favour five periodicities after the sixth full observing period and 197 velocities. This fifth signal at a period of 190 days did not disappear after the inclusion of another 43 velocity measurements and was clearly present in the data published by \citet{forveille2011}. However, another independent data set obtained using the HIRES did not support the existence of this signal. This was evident when the signal could not be detected significantly in the combined HARPS and HIRES data set. Therefore, we conclude that the 190 day periodicity in the HARPS data set, while being significantly present, was likely not of planetary origin but a spurious periodicity caused by some source of (systematic) noise that had variations resembling periodic behaviour. Alternatively, it is possible that the noise model does not describe the HIRES and/or HARPS data well enough, which might result in the inability to detect this signal in the combined set. However, we do not see any evidence of such inadequate modelling as the two data sets are consistent according to the Bayesian model inadequacy criterion of \citet{tuomi2011b}.

The combined HARPS and HIRES data set, however, did contain a fifth periodic signal. This time the weak fifth signal occurred at a period of 43.7 days. This result was obtained by modelling the two velocity sets using Gaussian white noise models that are almost a standard way of describing radial velocities in practice. As was shown recently by \citet{baluev2012}, the radial velocity noise of GJ 581 is not white but contains significant correlations (red noise) that, if not accounted for, could mimic periodic behaviour. We found this to be the case and observed that the statistical model containing a first order moving average component improved the model significantly and enabled us to describe the combined data the best with only four periodic signals. Therefore, the 43.7 day signal observed in the combined data set using the white noise model is the most likely a spurious signal caused by insufficient modelling and consequent misinterpretation of the data and is not a genuine Doppler signal of planetary origin. This does not, however, imply beyond reasonable doubt that the 43.7 day periodicity is not a genuine signal because it is possible that the noise model is still not optimal\footnote{Actually this is not only possible but rather certain because the origin, nature, time-evolution, and other features of radial velocity noise are very poorly understood at the moment.} and that a better model could enable the recovery of the signal.

These results show that not only can insufficient modelling lead to detections of false positives, but relying on a single data set can also lead to detections of signals that are not supported by other independent measurements. False positives can be caused by these two sources in practice and care is needed to assess whether any given weak signal in any single data set is caused by either of these two effects or not. However, our results suggest that while the Bayesian detection criteria are not particularly sensitive to false positives, they can fall a victim of these two sources of false positives.

Finally, based on our results, it looks like there are four and only four planets orbiting GJ 581. This means that the 32 day signal observed by \citet{vogt2012} was a false positive, though the evidence in favour of this signal in their analyses was moderate at most anyway. Also, the periodogram analyses of \citet{baluev2012} obtained a decreased significance for the companion GJ 581 d because of the inconsistensies in the periodogram analyses described in Section \ref{sec:periodograms} and in e.g. \citet{tuomi2012}. However, the HARPS velocities of \citet{forveille2011} obtained from the spectra using the cross-correlation function method are not optimal \citep{pepe2002} but their accuracy can be significantly improved when using more sophisticated spectral reduction methods, such as the TERRA algorithm \citep{anglada2012,anglada2012b}. Improving the noise modelling can also lead to more accurate descriptions of the velocities and therefore enable the detections of one or some of the signals that were only obtained from the HARPS data or with a particular noise model at 190 and 44 day periods. A better understanding of the noise inherent in measuring RVs for M-dwarf stars is essential because current \citep{barnes2012} and future \citep{ramsey2008,mahadevan2009} planet search surveys aim to focus their attention on the inactive and slowly rotating subset of these stars \citep[see][]{jenkins2009}. For these reasons, and because GJ 581 is still a target of e.g. the HARPS and HIRES exoplanet surveys (according to \citet{vogt2012} there is already a considerable amount of additional HIRES data), we do not expect that the debate over the number of $k$ for GJ 581 is over. This is especially interesting because the period-space between GJ 581 c and d is known to contain dynamically stable orbits for low-mass planets and happens to correspond to the liquid water habitable zone of the star.

\begin{acknowledgements}
M. Tuomi is supported by RoPACS (Rocky Planets Around Cool Stars), a Marie Curie Initial Training Network funded by the European Commission's Seventh Framework Programme. J. S. Jenkins acknowledges funding by Fondecyt through grant 3110004 and partial support from Centro de Astrof\'isica FONDAP 15010003, the GEMINI-CONICYT FUND and from the Comit\'e Mixto ESO-GOBIERNO DE CHILE.
\end{acknowledgements}


\end{document}